\documentclass{aa}
\usepackage{graphicx}

\usepackage{wasysym}

\newcommand{\Teff}{\mbox{$T_{\rm{eff}}$}}
\newcommand{\logg}{\mbox{log $g$}}
\newcommand{\feh}{\mbox{[Fe/H]}}

\newcommand{\ofe}{\mbox{[O/Fe]}}
\newcommand{\ie}{i.e.~\/}
\newcommand{\eg}{e.g.~\/}
\newcommand{\Hi}{H\,{\sc i}}

\newcommand{\Oi}{O\,{\sc i}}

\newcommand{\gae}{\lower 2pt \hbox{$\, \buildrel {\scriptstyle >}\over {\scriptstyle
\sim}\,$}}
\newcommand{\lae}{\lower 2pt \hbox{$\, \buildrel {\scriptstyle <}\over {\scriptstyle
\sim}\,$}}

\begin{document}

  \title{Neutral oxygen spectral line formation revisited
with new collisional data: large departures
from LTE at low metallicity}

  \author{D. Fabbian\inst{1,2},
     M. Asplund\inst{3},
     P.~S. Barklem\inst{4},
     M. Carlsson\inst{5},
     D. Kiselman\inst{6}}

  \offprints{D. Fabbian, \\
\email{damian@iac.es} }

  \institute{Research School of Astronomy \& Astrophysics, The Australian
  National University, Mount Stromlo Observatory, Cotter Road, Weston
  ACT 2611, Australia
        \and Current address: Instituto de Astrof\'isica de Canarias,
        Calle Via L\'actea s/n, E38205, La Laguna, Tenerife, Espa\~na
        \and
        Max Planck Institute for Astrophysics,
        Postfach 1317, 85741 Garching b. M\"unchen, Germany
        \and Department of Physics and Astronomy, Uppsala University,
        Box 515, 751-20 Uppsala, Sweden
        \and  Institute of Theoretical Astrophysics, University of
        Oslo, P.O. Box 1029, Blindern, N-0315 Oslo, Norway
\and The Institute for Solar Physics of the Royal Swedish Academy of
        Sciences, AlbaNova University Centre, 106 91 Stockholm,
        Sweden}

  \date{Received 2008 / Accepted 2008}

  \abstract
  {} 
  %
  %
  {A detailed study is presented,
  including estimates of the impact on elemental abundance analysis,
  of the non--Local Thermodynamic Equilibrium (non--LTE)
  formation of the high-excitation neutral oxygen $777$~nm triplet in model atmospheres
  representative of stars with spectral types F to K.}
  %
  %
  {We have applied the statistical equilibrium code {\small MULTI} to
  a number of plane--parallel {\small MARCS} atmospheric models
  covering late-type stars ($4500 \le T_{\rm eff} \le 6500$\,K, $2 \le \logg \le 5$\,[cgs], and
  $-3.5 \le \feh\, \le 0$). The atomic
  model employed includes, in
  particular, recent quantum-mechanical electron collision data.}
  %
  %
  {We confirm that the O\,{\sc i} triplet lines form
  under non--LTE conditions in late-type
  stars, suffering \protect{\it negative} abundance corrections with
  respect to LTE. At solar metallicity, the non--LTE effect,
  mainly attributed in previous studies to photon losses in the triplet
  itself, is also driven by an additional significant contribution from line
  opacity. At low metallicity, the very pronounced departures from LTE are
  due to overpopulation of the lower level (3s\,$^5$S$^{\rm o}$) of the
  transition. Large line opacity stems from triplet-quintet intersystem
  electron collisions, a form of coupling previously not considered or
  seriously underestimated.  The non--LTE effects become generally
  severe for models (both giants
  and dwarfs) with higher $T_{\rm eff}$.
  Interestingly, in metal-poor
  turn-off stars, the negative non--LTE abundance corrections
  tend to
  rapidly become more severe towards lower metallicity. When neglecting
  H collisions, they amount to as much as $|\Delta\log\,\epsilon_{\rm O}|
  \sim 0.9$\,dex and $\sim 1.2$\,dex, respectively at \mbox{[Fe/H]}$=-3$
  and \mbox{[Fe/H]}$=-3.5$. Even when such collisions are included, the LTE abundance remains a serious
  overestimate, correspondingly by $|\Delta\log\,\epsilon_{\rm O}|
  \sim 0.5$\,dex and $\sim 0.9$\,dex at such low metallicities. Although the
  poorly known inelastic hydrogen collisions thus remain an important
  uncertainty, the large metallicity-dependent non--LTE effects
  seem to point to a resulting ``low'' (compared to LTE) [O/Fe] in metal-poor halo stars.}

   \keywords{line: formation -- stars: abundances -- stars: late-type --
     Galaxy: evolution}

  \authorrunning{D. Fabbian et al.}
  \titlerunning{O\,{\sc i} $777$~nm triplet non--LTE line formation}

  \maketitle


\section{Introduction}

After the primordially ubiquitous H and He, oxygen is the third most
abundant chemical element in the Universe, and the first among those
produced via stellar nucleosynthesis. It is a catalyst in the CNO
process, which is the main conversion channel of H into He in the
interiors of stars more massive than the Sun. The oxygen content of
stars and its abundance ratio with other chemical elements derived
from observations of metal-poor halo stars are crucial (e.g.
Matteucci \& Fran{\c c}ois 1992) to constraining Galactic Chemical
Evolution (GCE) models. Oxygen, like the other $\alpha$-elements,
has been known for a long time (e.g. Conti et al.  1967) to show an
overabundance with respect to iron in stars of low metallicity. This
is expected since its main production site is in massive stars which
end their lives as Type II supernovae (SNe II, Arnett 1978; Clayton
2003). Due to the longer lifetimes of their precursors, the delayed
release of additional iron from type Ia SNe starts to affect
\ofe\footnote{By definition,
[A/X]=$\log(N_{A}/N_{X})_{*}-\log(N_{A}/N_{X})_{\odot}$}\, only at a
later stage. The oxygen abundance in unevolved very metal-poor stars
should therefore reflect the enrichment by the very first
generations of SNe.

The oxygen abundance at low metallicity has been the focus of a
plethora of studies. Still, a long-standing debate exists on the
exact [O/Fe] trend in metal-poor stars (see e.g. discussion in
Asplund 2005 and Mel\'{e}ndez et al. 2006 and references therein).
Here, we will only briefly mention some pertinent works related to
the subject.

Neutral oxygen lines used in abundance studies of solar-type stars
include the high excitation line at $615.8$~nm, the weak forbidden
[\Oi] lines at $630.0$~nm and $636.4$~nm, and the infrared triplet
at $777$~nm. The latter is the only strong oxygen atomic feature
observable in spectra of late-type stars. While it is known to
suffer from large non--LTE effects in early-type stars (Johnson,
Milkey \& Ramsey 1974; Baschek, Scholz \& Sedlmayr 1977), there is
little consensus on the estimated size of the abundance corrections
for cooler stars of solar-type. For example, the formation of the IR
triplet lines in the Sun has been the object of a number of
investigations. With a semiempirical approach, Altrock (1968) found
that the non--LTE source function drops below the Planck function in
the solar photosphere, a result later confirmed by the multilevel
non--LTE study by Sedlmayr (1974). More recently, the non--LTE
abundance corrections in the Sun have been estimated to be between
$\sim -0.1$ and $-0.25$~dex (Kiselman 1991, 1993; Gratton et al.
1999; Takeda 2003; Takeda \& Honda 2005; Asplund et al. 2004),
mostly depending on whether efficient collisions with neutral H
atoms are adopted or not. These authors explain the non--LTE line
strengthening as due to photon losses in the lines, while the
population (and thus the line opacity) of the lower level in the
$777$~nm transition is thought to stay close to LTE.

Non--LTE effects on oxygen lines in solar-type stars have been
reviewed by Kiselman (2001). Statistical equilibrium calculations
using a 16-level oxygen model atom (Kiselman 1991) have predicted an
enhancement towards low metallicity of the (negative) non--LTE
abundance corrections on the permitted $777$~nm \Oi\, lines,
suggesting that an LTE approach will seriously overestimate the
derived oxygen abundance. The [O/Fe] ratio derived in LTE
using this abundance indicator would thus be increasingly reduced in
metal-poor stars once non--LTE effects are considered. Other
calculations have instead found almost
\feh-independent non--LTE corrections (Takeda 2003).

Very high overabundances of oxygen compared to iron have
occasionally been derived at low metallicity. Analysing the \Oi\
777\,nm lines in metal-poor dwarfs, Abia \& Rebolo (1989) found a
steep monotonic increase in the oxygen-to-iron ratio for lower
metallicities, reaching \ofe\,$> 1$ below \feh\,$\sim -2$). Those
authors investigated non--LTE effects for the triplet lines but
claimed that they should be less severe than $-0.2$~dex, therefore
still leaving a residual increasing trend. Their high abundance
estimates were criticized in King (1993) as mainly due to too low
effective temperature estimates. Other authors (Gratton \& Ortolani
1986; Barbuy 1988; Sneden et al. 1991; Spite \& Spite 1991; Kraft et
al. 1992; Carretta, Gratton \& Sneden 2000; Cayrel et al. 2004;
Spite et al. 2005) found \ofe$\simeq 0.4-0.6$ in late-type stars
with \feh$\lae -1$, in particular using the [\Oi] lines in
metal-poor giants. This nearly plateau-like behaviour (with oxygen
content varying much more slowly than iron at low \feh, or even
remaining constant) could be interpreted as the result of the
contribution by type II SNe alone to the early Galactic
nucleosynthesis (see e.g. Matteucci \& Greggio 1986). While some
authors have reported a fair agreement between the near-IR $777$\,nm
oxygen lines and the forbidden line (Mishenina et al. 2000; Nissen
et al. 2002), others (Takeda 2003; Fulbright \& Johnson 2003)
derived higher (up to several tenths of a dex) abundance values from
the triplet. In particular -- even when applying to the observations
the non--LTE corrections derived in his own investigation -- Takeda
still derived apparently too high (by several tenths of a dex)
oxygen abundance from the IR triplet lines in metal-poor halo stars,
a possible signal of underestimated non--LTE effects. At
\feh\,$\apprle -3.5$, even more extreme examples of the oxygen
conflict have been found, with differences of up to $1.55$~dex
between [\Oi]-based and triplet abundances, and exceptionally large
\ofe$\apprge 2$ oxygen enhancements (e.g.  Depagne et al. 2002;
Israelian et al. 2004).

Unfortunately, the neutral atomic oxygen lines, in particular the
LTE-obeying [\Oi] line, tend to become vanishingly weak in
metal-poor solar-type stars. Thus, molecular bands of OH in the UV
(electronic transitions) and near IR (vibration-rotation lines) have
also been used. The former are relatively strong and thus detectable
in FGK stars down to low \feh, the latter instead require high S/N
and are only seen in spectra of cool stars (\Teff$<5000$~K). Results
from both are however affected by likely important 3D effects, in
particular in the case of the UV lines (Asplund \& Garcia Perez
2001), which may also suffer from the problem of missing opacity
(Balachandran \& Bell 1998; Asplund 2004) and possible issues
related to continuum location, blends in the spectra, atmospheric
extinction/cutoff and low CCD sensitivity. Various other
uncertainties -- \eg on oscillator strengths for both molecular
bands -- and incomplete knowledge of non--LTE effects on molecular
line formation may also be important. Generally higher values of
oxygen abundance (compared to those from the weak IR molecular
features) are derived from the OH lines in the UV (Balachandran,
Carr \& Carney 2001; Mel\'{e}ndez \& Barbuy 2002; Barbuy et al.
2003), but the above effects may indeed already for the most part
explain the large discrepancy. Other authors (Israelian, Garc{\'\i}a
L\'opez \& Rebolo 1998; Boesgaard et al. 1999) have instead found
reasonable agreement between the O abundances derived for a sample
of mainly turn-off stars from the two indicators. In any case, it
seems that consistent oxygen abundances using {\it all} the
different diagnostics, namely permitted, as well as forbidden and
molecular lines, can hardly be obtained at low metallicity.
Departures from homogeneity and from LTE have been at various times
invoked for solving the inconsistency. A combination of other
factors, including determination of stellar parameters/temperature
scale, quality of observations and of reduced data, adopted solar
abundances and so on may additionally be of concern and it is hard
to disentagle the different effects. To complicate the issue
further, one needs to consider that some authors have derived oxygen
abundances at low metallicity from unevolved stars, others from
giants, and it could be hypothesized that the lower oxygen values
usually found for the latter are due to a depletion process in their
atmospheres. However, Bessell, Sutherland \& Ruan (1991) discussed
evidence that oxygen abundances from OH were similar in dwarfs and
giants. More recently, Garc\'{\i}a P{\'e}rez et al. (2006) confirmed
a reasonable agreement between OH and [\Oi] abundances in metal-poor
subgiants, finding that \Oi\, triplet abundances give instead higher
values even when corrected for non--LTE effects. It is therefore
important to check if such residual discrepancy is linked to
problems in the previous modelling of the formation of the permitted
lines.

Here, we thus focus on investigating departures from LTE in
1D. Crucially, we explore the very low metallicity range, to
understand if the non--LTE abundance corrections depend strongly on
metallicity. In the next section, we describe our non--LTE
calculations, and give results in Section 3.  We then embark on a
comparison with other non--LTE studies of the
\Oi\,$777$\,nm absorption lines (Sect. 4). Finally, the last two
sections are dedicated to a discussion of the impact on the \ofe\,
ratio at low metallicity, and to some concluding remarks.

\section{Non--LTE calculations}

\subsection{Method}

The method employed is similar to that described in Fabbian et al.
(2006). In order to derive the non--LTE populations of the O atomic
levels and the strength of relevant spectral lines, it is necessary
to solve the coupled rate and radiative transfer equations
simultaneously. The code {\small MULTI} (Carlsson 1986) in its
version 2.3 was used to perform the statistical equilibrium
calculations for a number of {\small MARCS} model atmospheres
(Gustafsson et al. 1975; Asplund et al. 1997), under the assumption
that oxygen is a trace element and its departures from LTE will not
affect the atmospheric structure. The atmospheric models tested have
a microturbulence of $\xi=1$. A 54-level oxygen model atom
(described below) was employed. We assumed complete redistribution
in a Voigt profile for the formation of the lines.

\subsection{Atomic model}
\label{fabss:atom}


\begin{flushleft}

\begin{figure}
    \begin{center}
    \includegraphics[width=8cm,height=11.5cm]{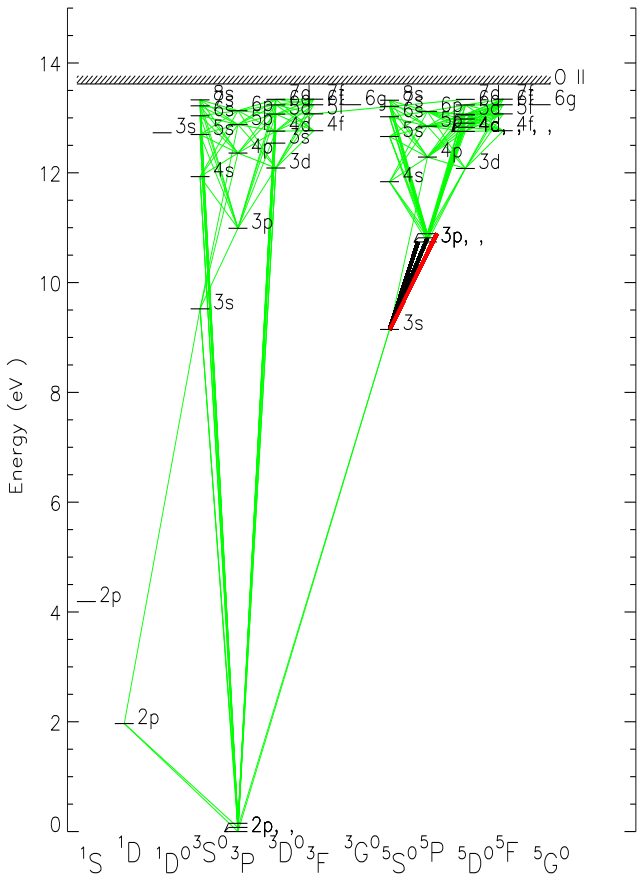}
    \caption{Grotrian diagram of the atomic model we employed, showing the
    53 energy levels of neutral oxygen and the single first ionization
    level. Lines connecting the various levels in the figure represent
    the 208 radiative bound-bound transitions included in the model. The
    $777$~nm transitions are marked with thicker
    lines\label{fabf:termdiag_oi_df54}}
\end{center}
\end{figure}

\end{flushleft}


We have constructed a model atom containing 54 energy levels and a
total of 258 radiative transitions (208 b-b and 50 b-f). The
Grotrian diagram of our atomic model is shown in
Fig.\,\ref{fabf:termdiag_oi_df54}. The necessary data for atomic
level energies and corresponding statistical weights were taken from
the NIST Atomic Spectra
Database\footnote{http://physics.nist.gov/PhysRefData/ASD/index.html}
version 3. Neutral levels are included up to an excitation
potential of 13.34 eV (i.e. $\sim 0.28$ eV below the continuum). The
model is complete to a principal quantum number of n$=7$.
Fine-splitting of energy levels was taken into account where
appropriate, in particular for the ground state and for the 3p $^5$P
level (upper level of the $777$\,nm \Oi\, triplet). For the
transitions of importance in solar work which were used in Asplund
et al. (2004), f-values are also from NIST while radiative and Stark
parameters come from VALD\footnote{Available at
http://ams.astro.univie.ac.at/vald/} (Vienna Atomic Line Database,
Piskunov et al. 1995 and successive updates). Data for bound-free
photoionization are from the Opacity
Project\footnote{http://vizier.u-strasbg.fr/topbase/topbase.html}
(Seaton 1987 and later updates). For collisional broadening due to H
atoms,
we adopt the classical Uns\"{o}ld approximation. Our tests reveal that
choosing an enhancement factor of two for the related damping constant
does not affect our results significantly. The contribution to
background opacity in the UV from lines of other elements was
accounted for by compiling NIST and VALD values. In general, the data
in the model atoms\footnote{Respectively, a 45- and a 23-level
atomic model} of Carlsson \& Judge (1993, hereafter CJ93) and
Kiselman (1993, hereafter K93) were used (see those papers for
description of the data sources) unless otherwise stated.

We have also included the most recent data for electron-impact
excitation, available up to level 2p$^3$ 4f $^3$F from
quantum-mechanical calculations (Barklem 2007a), a
novel feature of our model atom which has proven one of the essential
improvements with respect to other studies. In particular, this has
allowed us to include accurate rates for transitions between
radiatively forbidden transitions due to electron collisions.

Since the estimated cross-sections were derived using LS-coupling,
they are null only for singlet-quintet transitions (total electronic
spin conservation rule) and for collisionally forbidden S$^{\rm e}$ --
S$^{\rm o}$ transitions. In broad terms, the collisional
cross-sections compare fairly well with available empirical data
(within the experimental error bars, see Fig. 2 in Barklem
2007a). Although admittedly still subject to uncertainties due to \eg
pseudo-resonances at higher energies, they should be more realistic
than previous estimates. They are in general comparatively larger than
H collisions by a few orders of magnitude,
except for fine-splitting levels\footnote{We account for fine
splitting also for electron-impact data, by using Barklem's
calculations (which originally refer to grouped levels) according to
conservation of total rates}. We stress here that the cross-sections
for electron-impact collisions in the radiatively forbidden transition
3s $^3$S$^{\rm o}$ -- 3s $^5$S$^{\rm o}$ are among the largest in the
calculations of Barklem (2007a). In particular (see Fig. 1 in that
work) they are larger than the other values for transitions arising
from the lowest levels (and between the first seven energy levels in
our atomic model).

Due to the deficiency of metals, and thus the decreased number of
free electrons in their atmosphere, rates due to collisions with
neutral H atoms may be particularly important in low-metallicity
stars. The approximation we adopted for excitation and ionization
via inelastic collisions with \Hi\, is based on recipes by Steenbock
\& Holweger (1984), which further generalize results in Drawin
(1968, 1969), where Thomson's classical theory for electron-atom
encounters was applied for collisions between identical particles.
The approach is admittedly less than ideal, but unavoidable due to
the lack of relevant experimental data and theoretical calculations
(see e.g. the discussion in Asplund 2005). In fact, quantum
mechanical calculations of rate coefficients exist for only very few
electronic transitions of oxygen atoms induced by collisions with H
(Krems, Jamieson \& Dalgarno 2006; Abrahamsson et al. 2007), and
they are unfortunately often of limited applicability for stellar
work due to the low range of temperatures explored.

Thus, in order to estimate the size of the main uncertainties in the
calculations, we have carried out, similarly to what was done in
Fabbian et al. (2006), different non--LTE calculations with a scaling
factor S$_{\rm H}$ for transitions between all levels, including for
radiatively forbidden transitions (by adopting a minimum f-value of
$10^{-3}$ in that case).

\vspace{0.7cm}

\section{Results}

\subsection{Non--LTE mechanisms}
\label{fabss:mech}


\begin{flushleft}

\begin{figure}
    \begin{center}
    \includegraphics[width=9cm,height=12cm]{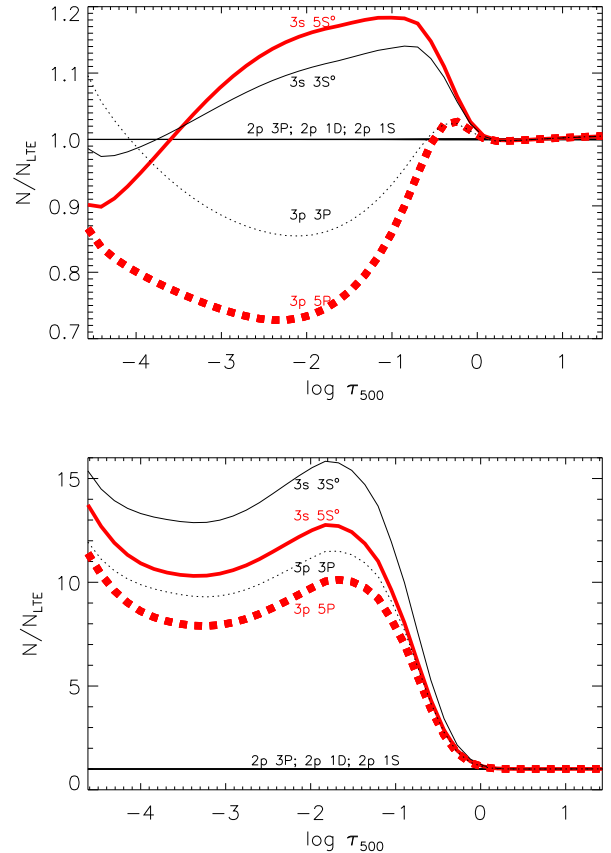}
    \caption{Departure coefficients in the level population of the
    lowest energy levels (up to 3p $^3$P) in our oxygen model atom (for
    the case without H collisions) are shown respectively for the Sun
    (upper panel) and a metal-poor turn-off
    atmospheric model with \feh$=-3$ (lower panel). Note the very different scales for
    the y-axes in the two panels. The thicker lines indicate the departure coefficients
    for lower and upper levels (solid and dashed curve respectively) in the $777$~nm
    triplet. Note that only one curve is plotted for the latter levels, because the small
    differences in the departure coefficients among those fine structure levels does not
    play an important role here\label{fabf:betas}}
\end{center}
\end{figure}

\end{flushleft}


The behaviour of the departure coefficients $\beta_i$ (defined as the
ratio of the populations in non--LTE and LTE, $\beta_i=n_i/n_i^{\rm
LTE}$) with atmospheric depth for various energy levels in the atomic
model is shown in Fig.\,\ref{fabf:betas} for the Sun and a metal-poor
turn-off model.

Non--LTE effects for the \Oi\, $777$\,nm lines at solar metallicity
are mainly caused by source function dilution, in the atmospheric
layers where the lines form, with respect to the Planck function that
represents the LTE expectation. This is due to radiative losses in the
triplet itself causing line strengthening compared to LTE. As seen in
Fig.\,\ref{fabf:betas}, the upper level (3p $^5$P) of the transition
gets underpopulated, while the lower level (3s $^5$S$^{\rm o}$) gets
overpopulated. If stimulated emission is neglected, this driving
effect can be well described as $S_l/B=\beta_u/\beta_l$. The two-level
approximation will give a reasonable description of the non--LTE line
formation in the Sun, implying that it is the processes in the
transition itself that matter most.

At low metallicity instead, particularly at the highest temperatures
explored, that approximation breaks down. The upper level (3s $^3$S$^{\rm o}$) of the
$130$~nm \Oi\, resonance lines is in effect metastable, since radiative de-excitation
will be immediately followed by an excitation (due to large
opacity in those lines). Thus, very large overpopulation compared to LTE develops in the
level, because of the larger net rate for it than in LTE. This is
caused by radiative transitions having this as their lower level (i.e. by
photon losses in higher excitation levels), combined with the low rate
coefficients out of the level.

The 3s $^3$S$^{\rm o}$ overpopulation tends to propagate to the 3s
$^5$S$^{\rm o}$ level via efficient intersystem collisional coupling,
and line strengthening of the $777$~nm lines results, due to much
increased line opacity. Given the large energy difference with the
ground state, collisions to it from the levels of interest are
inefficient in maintaining LTE. The dominant contribution to the
non--LTE effect at low metallicity thus comes from line opacity,
because the lower level of the IR triplet is strongly overpopulated
with respect to LTE (Fig.\,\ref{fabf:betas}). The non--LTE abundance
corrections then become $\sim -0.5$ dex at
\feh$=-3$ for a typical metal-poor turn-off halo star, and up to
$-0.9$~dex if neglecting H collisions.

\begin{flushleft}
\begin{figure*}
    \begin{center}
    \includegraphics[width=15cm,height=8cm]{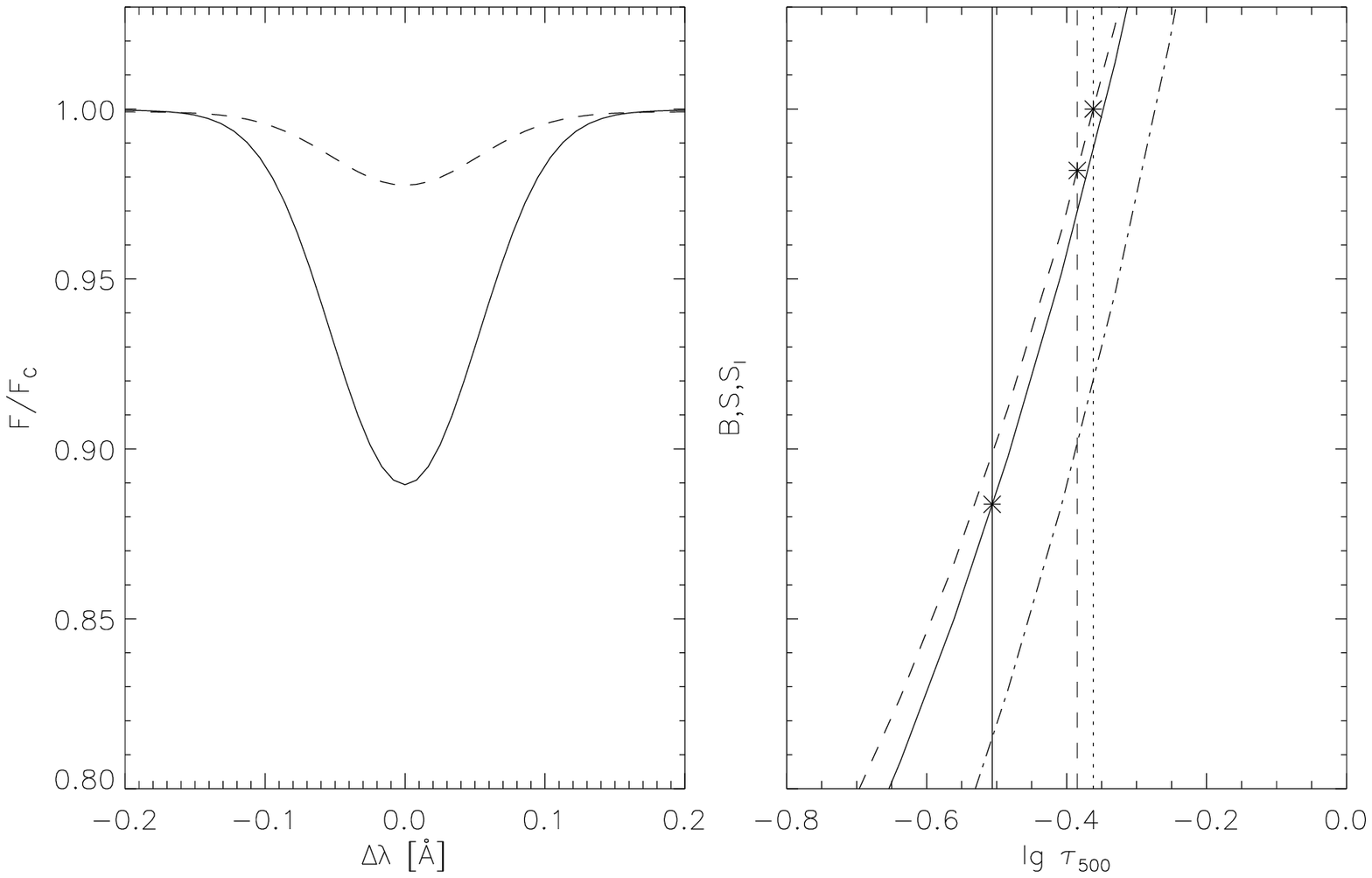}
    \includegraphics[width=15cm,height=8cm]{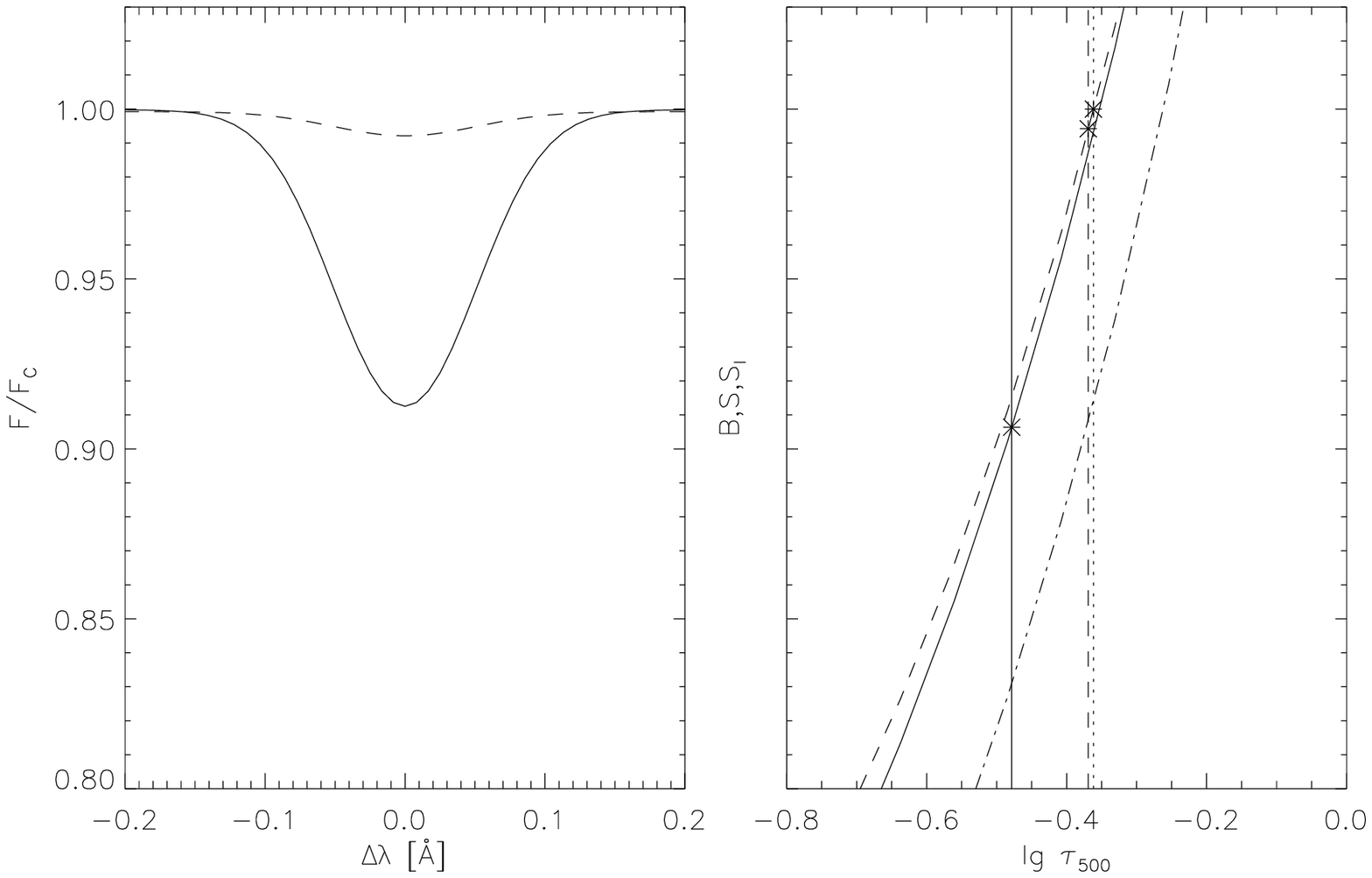}
    \caption{The various panels refer to the oxygen feature at
    $777.2$~nm\ (the bluest and strongest of the \Oi\, triplet lines),
    for \mbox{[Fe/H]}$=-3.00$ (top panels) and $-3.50$ (bottom
    panels). The atmospheric models have
    \mbox{$T_{\rm{eff}}$}$=6500$~K, \mbox{log $g$}$=4.00$ [cgs] and
    solar-scaled, alpha-enhanced ([O/Fe]$=+0.4$) oxygen abundance. The
    case with no hydrogen collisions is shown here. {\it Top}: the
    resulting LTE (dashed curve) and non--LTE (solid curve) flux
    profiles, normalized to continuum flux, for the spectral feature
    considered are shown in the left panel, while normalized values of
    B (dashed), S (solid) and S$_l$ (dash-dot) are shown in the right
    panel. The vertical scale in the two cases is therefore the same.
    The left panel illustrates how at low metallicity the line
    becomes much stronger in non--LTE compared to LTE. In the right
    panel, vertical lines show the continuum (dotted), LTE line center
    (dashed) and non--LTE line center (solid) Eddington-Barbier values
    of $\tau_{\nu}$ for which the continuum flux corresponds to the Planck
    function at that depth
    (i.e. F$^{cont}_{\nu}$=$\pi$S$_{\nu}$($\tau^{cont}_{\nu}$=E.B.)). The
    corresponding starred symbols on the various curves therefore
    represent the Eddington-Barbier formation height and flux,
    respectively for continuum, LTE line center and non--LTE line
    center. Note that the Eddington-Barbier value is
    $\tau^{cont}_{\nu}=2/3$ exactly when the source function varies
    linearly with optical depth, while it is slightly smaller in this
    case. The right panel shows that at low metallicity the source
    function effect is small. {\it Bottom}: as above, but for
     \mbox{[Fe/H]}$=-3.50$ \label{fabf:mc_fig}}
\end{center}
\end{figure*}
\end{flushleft}

As seen in Fig.\,\ref{fabf:mc_fig}, in non--LTE the line formation at
very low metallicity is shifted to higher atmospheric layers. Since
the population of the lower level in the oxygen $777$\,nm triplet
transition rises very steeply with respect to the LTE expectation when
moving outward in the atmosphere, the $\beta$ departure coefficient
will become large. Fig.\,\ref{fabf:mc_fig} also shows that the effect
(on line-center flux) of photon losses in the line, causing the source
function to drop below the Planck function, is small at low
\feh. The opacity effect dominates, with optical depth unity moving up
in the atmosphere because of the overpopulation of the lower level of
the transition. The large line opacity at the shallower depths where line
formation occurs in non--LTE will thus make the absorption stronger.


Even with decreased free electrons at low metallicity, sufficient
intersystem collisional coupling is maintained between levels of
similar excitation in the triplet and quintet systems. The 3s
$^3$S$^{\rm o}$ and the 3s $^5$S$^{\rm o}$ levels are in fact very
close energetically and, despite the corresponding transition being
radiatively forbidden, they are strongly coupled via collisions, which
explains why their departure coefficients behave similarly, as seen in
Fig.\,\ref{fabf:betas}. Due to decreased number of electrons and to
small rates for collisional transitions with larger energy separation,
electron collisions are more and more ineffective in driving
the line formation closer to LTE via depopulating these two levels via
other channels, e.g. to the ground state. The less efficient
thermalization via impacts is thus unable to balance the tendency of
the triplet levels to overpopulate. In addition, the lower level of
the $777$~nm transition is effectively metastable due to negligible
net radiative rates to the ground state.

Among the IR \Oi\, triplet lines, the strongest (bluest) suffers
generally more severe non--LTE effects. This is due to it having
relatively more important photon losses, and given that it will form
further out where departures from LTE are largest. The difference in
the estimated abundance correction among the $777$~nm lines usually
remains within $0.05$~dex, but grows towards solar metallicity. The
abundance corrections differ by as much as $\sim -0.15$~dex at
\Teff$=6500$~K, \logg$=2$, \feh$=0$, where very large non--LTE
effects are found. Thus, adopting the same correction for all three
lines is not appropriate, especially in such case, and detailed
non--LTE calculations are necessary. This is due to substantially
different line formation depths, giving larger non--LTE effects for
the line that is formed further out, i.e. the strongest (bluest)
one, because of outward increase of the departure coefficients.

Across all of the parameter space explored, the upper level (3s
$^3$S$^{\rm o}$) in the \Oi\, UV resonance lines tends to get
overpopulated. Large radiation excess is maintained in the resonance
lines, becoming particularly strong at low metallicity, for high
gravity models.

In the $777$~nm lines instead, the ratio of J over B tends to be $<
1$ due to photon losses. This is relatively more important in stars
like the Sun, and causes the upper levels in the $777$~nm triplet to
depopulate.

Photoionization was found not to be important across the parameter
space explored in this study.

\vspace{0.7cm}

\subsubsection{Sensitivity of line formation: main atomic transitions}


\begin{flushleft}

\begin{table}
 \centering
 \caption{Results (for the strongest triplet line at $777.2$~nm)
 obtained using our 54-level atomic model.  The non--LTE and LTE
 equivalent widths are given to show
 the sensitivity to various mechanisms.
 The changes {\it with respect to the standard case}
 (\Teff$=6500$~K, \logg$=4$, \feh$=-3$, $\xi=$1,
 $\log\,\epsilon_{\rm O}=6.5$, and including the collisional data of
 Barklem 2007a)
 are: decrease in metallicity by half dex (to \feh$=-3.5$); and
 decrease in oxygen abundance by half dex (to $\log\,\epsilon_{\rm O}=6.00$). For the
 case including H collisions, additional tests are, following the
 order listed: removing the $130$~nm \Oi\, resonance lines; removing
 also the intersystem $136$~nm UV \Oi\, lines; removing all of the
 new electron collisions by Barklem except in the radiatively forbidden
 transition between
 levels 3s $^3$S$^{\rm o}$ and 3s $^5$S$^{\rm o}$; including all
 the new electron collisions by Barklem except the previous one;
 no 3s $^3$S$^{\rm o}$
 and 3s $^5$S$^{\rm o}$ coupling (i.e. removing both electron and H
 collisions between those levels); use of old electron collision
 data from 23-level model atom of K93 instead of Barklem's data; same
 as previous but with new
 electron collision data between levels 3s $^3$S$^{\rm o}$ and 3s
 $^5$S$^{\rm o}$; old electron data but no H collision coupling between the
 latter levels;
 neglect of UV background line opacities; exclusion of the opacity
 contribution by the blending H Ly-$\beta$ line; and decrease by
 $1$~dex in Si abundance\label{fabt:df54}}
 \begin{tabular}{l|cc}
        & W$_{\rm non-LTE}$ & W$_{\rm LTE}$ \\
        & [m\AA]            & [m\AA]        \\
  \hline \hline
{\bf S$_{\bf{H}}\bf=0$}  &                   &   \\
        &   &           \\
standard case & 14.2   & 2.9   \\
\feh$=-3.5$  & 10.9   & 1.0   \\
$\log\,\epsilon_{\rm O}=6.0$  &  6.3   & 1.0   \\
 \hline
{\bf S$_{\bf{H}}\bf=1$}  &    &           \\
        &    &           \\
 standard case        &  6.7 & 2.9 \\
 \feh$=-3.5$         &  4.2 & 1.0 \\
 $\log\,\epsilon_{\rm O}=6.0$         &  2.9 & 1.0 \\
 no $130$~nm \Oi\, lines\,                  &  3.9 & 2.9 \\
 no $130$~nm, no $136$~nm \Oi\, lines                   &  3.6 & 2.9 \\
 3s -- 3s Barklem e$^-$                       &  7.2 & 2.9 \\
 all Barklem e$^-$ but 3s -- 3s               &  6.5 & 2.9 \\
 no 3s -- 3s coupling                 &  5.2 & 2.9 \\
 old e$^-$ collisions                      &  6.2 & 2.9 \\
 old e$^-$ collisions, 3s -- 3s Barklem e$^-$ &  6.4 & 2.9 \\
 old e$^-$ collisions, no O+H 3s -- 3s  &  5.0 & 2.9 \\
 no UV background line opacity       &  7.8 & 2.9 \\
 no H Ly-$\beta$ opacity                &  6.8 & 2.9 \\
 Si*0.1                               &  9.4 & 2.9 \\
 \hline
 \hline
 \end{tabular}
\end{table}

\end{flushleft}



\begin{flushleft}

\begin{figure*}
    \begin{center}
    \caption{
    (only available online)
    Results from multi-{\small MULTI} runs
    for a solar model, and for a star with
    \mbox{$T_{\rm{eff}}$}$=6500$~K, \mbox{log $g$}$=4.00$ and
    \mbox{[Fe/H]}$=-3$, when neglecting H collisions.
    \label{fabf:multiMULTI_noH}}
\end{center}
\end{figure*}

\end{flushleft}


In Table \ref{fabt:df54} we provide the equivalent widths resulting
from some of the various tests carried out on our 54-level atomic
model in order to check the sensitivity of the $777$~nm triplet line
formation to various transitions and processes in the atomic model.
The {\it standard case} shown (\Teff$=6500$~K, \logg$=4$, \feh$=-3$,
$\xi=$1, $\log\,\epsilon_{\rm O}=6.5$) includes the new
electron-impact data by Barklem (2007a). It results in
$\Delta\log\,\epsilon_{\rm O} \sim -0.9$ and $\sim -0.5$~dex without
and with the inclusion of H collisions respectively. The
corresponding non--LTE abundance corrections at \feh$=-3.5$ increase
to $\Delta\log\,\epsilon_{\rm O} \sim -1.2$ and $\sim -0.9$~dex
respectively. The modifications were performed on this {\it standard
case}, as listed in Table \ref{fabt:df54}. From the results there
shown, it is clear that the most important transitions are the \Oi\,
resonance lines (together with the corresponding UV background
opacity), and the intersystem collisional coupling between the 3s
$^3$S$^{\rm o}$ and 3s $^5$S$^{\rm o}$ states. The model atom
employed in the {\it standard case} is also the one we then used to
derive our non--LTE corrections across the whole parameter space
(see Section\,\ref{fabss:paramspace}).

We also carried out ``multi-{\small MULTI}'' tests, with results for
the case without H collisions shown in
Fig.\,\ref{fabf:multiMULTI_noH} (only available online). Through
such tests, the relative effect on the O\,{\sc i}\, $777$~nm triplet
line strength of multiplying each of the different rates for
radiative and electron collisional transitions individually by a
factor of two is revealed. One can thus grasp the impact of
different atomic transitions on the formation of the lines of
interest. As seen in Fig.\,\ref{fabf:multiMULTI_noH}, and as found
by previous authors (e.g. Kiselman 1991, 1993), the formation of the
O\,{\sc i}\, triplet in the Sun is mostly influenced by photon
losses in those lines and can be described reasonably well by a
two-level approximation. We stress however that our calculations
also show an additional contribution from line opacity, due to the
lower level of the triplet transition getting overpopulated. This
increases the resulting non--LTE effects for the Sun to a larger
amount than usually adopted, up to $\sim -0.3$~dex when neglecting H
collisions.

At these solar or moderately low metallicities, the line opacity
effect in the
\Oi\, IR triplet starts to dominate (over that caused by the source
function drop) for hot giants. It becomes much more predominant at
very low metallicity, where we find an increase in the size of the
abundance corrections with increasing
\Teff\, and decreasing \feh, related, respectively, to increased
level overpopulation via collisions, and to decreasing continuum
opacity. An increase in gravity at metallicities below \feh$\sim -2$
will also tend to cause larger non--LTE corrections. This effect is
related to the increased sensitivity of the line formation process to
intersystem collisional coupling, in that regime.

So, while in the Sun both source function drop and line opacity are
important, with the first being predominant, the metal-poor turn-off
model is mostly affected by the latter. The effect of the line
source function dilution does not change significantly with \feh\,
at very low metallicity, its contribution to line strengthening
remaining $\lae 10\%$. For parameters typical of metal-poor turn-off
stars, we find a trend of increasing non--LTE corrections with
decreasing \feh. This is caused by the behaviour of continuous
opacity at the UV wavelengths of interest (i.e., around the \Oi\,
$130$~nm features), which decreases substantially (up to $\sim
15$\%) across the line formation layers when the metal content goes
from \feh$=-3$ to $-3.5$. Even though the contribution of Rayleigh
scattering to total continuous opacity is dominant for layers around
and above those where the $777$~nm features are formed, it does not
vary when metallicity decreases from \feh$=-3$ to $-3.5$. Instead,
it is the decreased contribution by Si photoionization to total
continuous opacity when [Fe/H] decreases (see Fig.\,\ref{fabf:out})
that deprives the photons in the \Oi\, resonance lines of an
alternative route. Their destruction probability via the
photoionization channel therefore decreases. As a consequence,
overpopulation (driven by photon losses in higher-excitation lines)
of the upper level in the \Oi\, resonance lines tends to increase.
At the same time, the lower level of the $777$~nm transition will
too correspondingly get more overpopulated due to efficient
intersystem collisional coupling. At low metallicity, the non--LTE
effect on the \Oi\, $777$~nm triplet feeds on the flow via
collisions from the 3s $^3$S$^{\rm o}$ state. The level occupation
numbers in the lower level of the IR oxygen triplet thus stay
closely coupled to the large overpopulation in the upper level of
the UV $130$~nm resonance lines, a situation which LTE cannot
predict.

At \feh$\sim -3$, the triplet lines become stronger when increasing
the radiative rates for the resonance lines, because that causes
larger overpopulation in level 3s $^3$S$^{\rm o}$, and in the lower
level of the $777$~nm triplet as a consequence of collisional
coupling. Note that in the figure, we evaluate the net flow
(indicated by the direction of the arrows for each transition) where
$\tau=1$ at line center for the $777$~nm triplet. This generally
gives a good indication of the driving mechanisms.  However, at low
metallicity, one can note that the radiative rate in the triplet is
upward. A check reveals that in this case the rates actually change
sign across the atmosphere. While at $\tau=1.0$ at line center there
is a small upward flow, the net rate turns to being downward in deep
layers. This last effect dominates, explaining the overall line
strengthening, as marked in the corresponding figure, when
increasing the radiative rate in the \Oi\, IR triplet. Excluding the
transition itself, the most significant impact on the triplet lines
at low metallicity is from pumping in the \Oi\, $130$~nm resonance
lines.

Concerning electron collisions, the main effects at low metallicity
are strengthening of the line through relatively efficient
triplet-quintet intersystem coupling via electron
collisions\footnote{As seen in Fig.\,\ref{fabf:multiMULTI_noH}, this
channel gives instead a weakening effect in the Sun. The reason is
clear from the previously discussed Fig.\,\ref{fabf:betas} which shows
that in the Sun, the departure coefficient is larger for 3s
$^5$S$^{\rm o}$ (lower level of $777$~nm transition, in the quintet
system) than for 3s $^3$S$^{\rm o}$ (upper level of \Oi\, $130$~nm UV
resonance lines, in the triplet system). An increase of this
collisional coupling channel in the Sun will therefore {\it decrease}
the overpopulation and thus the opacity effect for the $777$~nm
transitions, making those lines form deeper in, where smaller non--LTE
effects are felt.} between the similarly highly-excited levels 3s
$^3$S$^{\rm o}$ and 3s $^5$S$^{\rm o}$ - which helps to propagate the
overpopulation from the former state to the IR oxygen triplet - and
line weakening via collisions to the ground state. At low metallicity,
both lower (3s $^5$S$^{\rm o}$) and upper (3p $^5$P) levels of the
triplet lines get very overpopulated, but by different amounts.

\subsubsection {Importance of \Oi\, UV transitions and background
line opacities}
\label{fabsss:UVopacity}

According to our calculations, even in metal-poor stars, at least
for frequencies near their quasi-saturated core, the extremely
optically thick $130$\,nm \Oi\, resonance lines (2p$^{4}$
$^{3}$P$_{2,1,0}$-3s 3S$^{o}$) form in the very topmost layers of
the photosphere as strong absorption features. Despite this,
their thermalization depth lies much deeper in than where
monochromatic optical depth in the lines is unity, due to the very
large amount of scattering. In particular their J/B$>>1$ where the
$777$~nm lines are forming. The properties of scattered photons are
decoupled from local conditions, which gives a typical non--LTE
situation.

The fact that the \Oi\, resonance lines are so optically thick even at
low [Fe/H] means that at the formation layers of interest, collisional
de-excitation from their upper (metastable) level to the lower level
of the $777$~nm transition will be favoured with respect to
de-excitation to the ground. This situation causes large non--LTE
effects at low metallicity. When removing the UV \Oi\, resonance lines,
the non--LTE effects remain important (see Table
\ref{fabt:df54}) due to photon losses from higher levels.

Previous studies (e.g., K93; Gratton et al. 1999) found that UV
radiation should not play a major role in the formation of the IR
oxygen triplet in very-metal poor star. They were, however, lacking
the detailed quantum mechanical calculations for intersystem
electron collisions which we employ here and which cause increased
non--LTE effects.

Background absorption from UV lines of other elements mitigates (by
providing an escape route other than collisional de-excitation and
collisional transfer to the quintet system) but does not extinguish
this effect. The exact amount of bound-free absorption from
different metals is however still debated (e.g. Balachandran \& Bell
1998; Asplund 2004), due to the possibility that continuum rates,
e.g.  for Fe I, may be larger than what usually adopted. As noted by
Allende Prieto, Hubeny \& Lambert (2003), predicting a realistic
flux in the UV is more complicated than for visible or IR
wavelengths. The knowledge of accurate bound-bound and
photoionization cross-sections and thresholds for metals becomes
important, with a large quantity of accurate atomic data required
for modelling. Below $\sim 1700$ \AA, Si I bound-free absorption
plays an important role. Rayleigh scattering by hydrogen atoms is
also important in the ultraviolet for cool stars, in particular at
low metallicity. In addition to bound-free metal edges, a sharp
increase in opacity is expected at short UV wavelengths also due to
the haze caused by background line absorption. Neglecting these
effects will seriously overestimate the UV radiation field within
the upper atmospheric layers of late-type stars.


\begin{flushleft}

\begin{figure}
    \begin{center}
    \includegraphics[width=9cm,height=12cm]{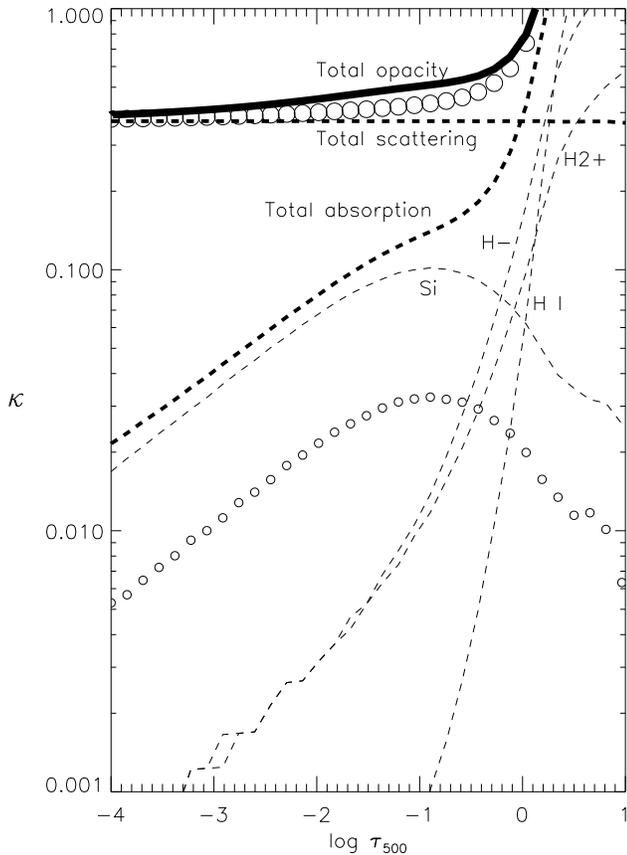}
    \caption{Contributions to total continuous opacity around the $130$~nm \Oi\,
    lines for a \Teff$=6500$, \logg$=4$, \feh$=-3$ model atmosphere.
    The absorption and the scattering part of total opacity are shown by the
    two thick
    dashed curves, as marked.
    For absorption, the importance of its main components
    (Si, H$^-$, H$_{2}^+$ and
    H I) is depicted. Total continuous opacity itself is shown by the thick solid curve. For
    comparison, we also show (small circles) the reduced contribution of Si at
    \feh$=-3.5$ and consequent reduced total continuous opacity (large circles). The
    rest of the sources of opacity in the figure do not change significantly with
    this decrease in metallicity\label{fabf:out}}
\end{center}
\end{figure}

\end{flushleft}


Fig.\,\ref{fabf:out} shows the main contributions to total
continuous opacity for UV wavelengths around the \Oi\, resonance
lines and a {\small MARCS} model having \Teff$=6500$~K, log g$=4$,
\feh$=-3$ and log O$=6.5$. As seen, scattering is roughly constant
across the atmospheric model. Its contribution to total opacity
dominates down to the atmospheric layers ($\log\,\tau_{500} \sim
-0.5$) where the IR oxygen triplet lines form.  The contribution of
absorption, even though smaller, is still significant, at such
layers predominantly thanks to Si\,{\sc i}\, bound-free edges. It is
this component which proves important to understand the trends of
the non--LTE corrections at low metallicity, since it is the only
one which varies significantly with \feh. At \feh$=-3.5$, its
contribution to absorption for the atmospheric layers of interest
reduces to $\sim 35\%$ of the value at \feh$=-3$. Thus, total
continuous opacity becomes significantly smaller. No other metal
component plays a significant role. Deeper in, absorption takes
over, due to H$^-$ and H$_{2}^+$ at first, and then to the rapid
increase of neutral hydrogen photoionization.

The larger continuum absorption at [Fe/H]$=-3$ compared to
[Fe/H]$=-3.5$ reduces the overpopulation in the upper state of the
UV \Oi\, resonance transitions because it provides an alternative
route to photons in those lines (absorption by the continuum instead
of by the line itself).

The non--LTE corrections are found to be very sensitive to the UV
radiation field. Thus, background opacity caused by lines of other
elements has been accounted for in the case of important
transitions, in particular the contribution of H Ly-$\beta$ and of
features around the \Oi\, $130$~nm resonance lines. The latter
inclusion has proven particularly important because it significantly
reduces the otherwise even more extreme non--LTE effects (see Table
\ref{fabt:df54}). In total, 151 lines from VALD were used, around
the 130, 103 and 136\,nm transitions\footnote{Respectively
accounting for, in order of decreasing importance, the effect of
features around the resonance lines, of absorption due to H
Ly-$\beta$, and of lines falling close to the intersystem oxygen
lines}. For these background features, the line source function was
assumed to be equal to the Planck function. This approximation is
sufficient also to describe the case of pumping by hydrogen
Ly-$\beta$ radiation. This was tested by solving the radiation
transfer problem for the hydrogen atom and taking the J field as
input for the non--LTE calculations on oxygen. The results showed
only very small differences with respect to our approximation, for
the formation of the oxygen triplet lines of interest here. Note
that H Ly-$\beta$ gives the single largest contribution to
background line opacity, however it is the total contribution of the
several blending metal lines falling around $130$~nm which is most
important, in particular due to S\,{\sc i}\, $130.23$~nm, which
falls close to line center of the bluest \Oi\, resonance line,
Si\,{\sc ii} $130.44$~nm, which is located in the blue wing of the
central resonance line but also affects the $103.22$~nm \Oi\, line
by depressing the continuum, and S\,{\sc i} $130.59$~nm, which falls
close to the center of the reddest resonance line. In addition, a
number of lines of Fe, Ca, P and other elements give a smaller
contribution to line haze. The lines around the \Oi\, UV intersystem
lines are instead not important.

Our study reveals that the inclusion of the UV opacity contribution
does not have a significant impact on the $777$~nm triplet lines in
the Sun, where the non--LTE effect is in that case at least in good
part controlled by processes in those same IR lines.

\subsubsection{Effects of collisions with electrons and hydrogen}
\label{fabsss:coll}


\begin{flushleft}

\begin{figure}
    \begin{center}
    \includegraphics[width=9.5cm]{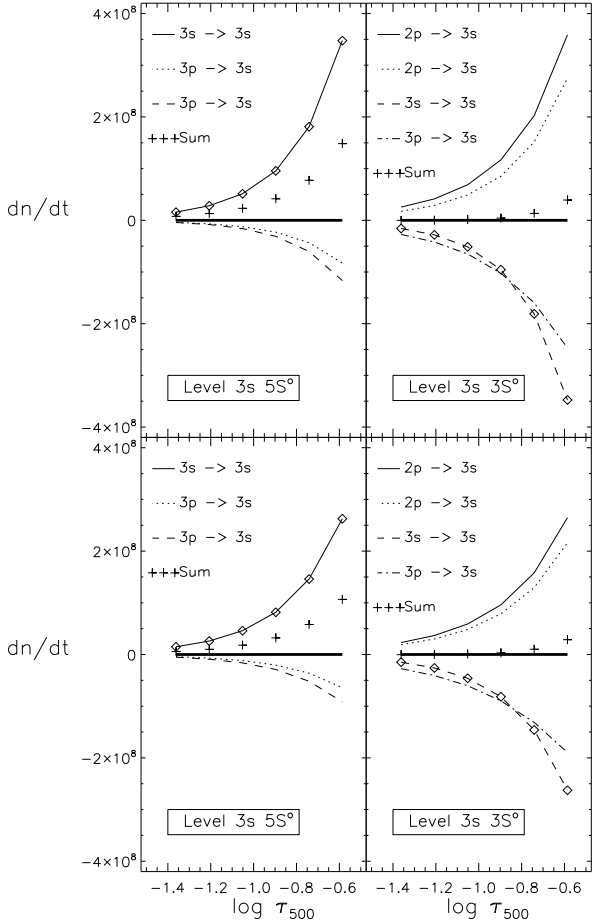}
    \caption{Net rates contributing to the balance of levels 3s
    $^5$S$^{\rm o}$ ({\it left}) and 3s $^3$S$^{\rm o}$ ({\it
    right}). Lines with and without diamond symbols respectively
    indicate the contribution from collisions and radiation to a given
    rate. Only the rates with large influence on the level population
    are shown (their summed contribution is represented by
    crosses). The thick horizontal line at dn/dt$=0$ marks the
    statistical equilibrium expectation that the total sum of {\it
    all} rates to and from a given level is equal to zero. The
    atmospheric model has \Teff$=6500$, \logg$=4$ and
    $\log\,\epsilon_{\rm O}=6.5$. Plots are for \feh$=-3$ (upper
    panels) and $-3.5$ (lower panels)
    respectively\label{fabf:levels5and6_feh}}
\end{center}
\end{figure}

\end{flushleft}


Figure \ref{fabf:levels5and6_feh} shows the net rates at low metallicity for
the high-excitation levels 3s $^5$S$^{\rm o}$ and 3s $^3$S$^{\rm
o}$. The figure clearly illustrates how the balance in the population
of the lower level of the $777$~nm oxygen triplet is mainly determined
by incoming flow due to collisional coupling - which transmits the
overpopulation present in the 3s $^3$S$^{\rm o}$ state - moderated by
radiative excitation in the IR lines.

We have included in our model atom a homogeneous set of
electron-impact excitation data, thanks to the availability of recent
theoretical quantum mechanical calculations (Barklem 2007a). Table
\ref{fabt:df54} shows the results of testing various modifications to
the atomic model, including for electron collisions. The effect of the
new electron-impact data is to give larger non--LTE corrections due to
increased intersystem coupling, in particular, in the 3s $^3$S$^{\rm
o}$ -- 3s $^5$S$^{\rm o}$ transition. This is an efficient channel due
to its large cross-section (it is a spin-flip transition induced by an
exchange interaction of the electrons) and its small ($\sim 0.38$~eV)
energy separation (thus low threshold).

Inelastic collisions with neutral H atoms as modelled here can play
a crucial role at low metallicity, where the scarcity of free
electrons means that rates due to H collisions can become important
in late-type stars due to the large H population density. Under the
assumption that van Regemorter's and Drawin's formulae well
reproduce the thermalization due to electron and H collisions
respectively, one derives that the ratio of the rates due to the two
processes is a function of the energy separation between the levels
involved and of temperature (Asplund 2005). These collisions can
thus become particularly efficient for levels that are very close in
energy. The approximate recipe available is based on work by Drawin
(1968, 1969) which generalizes the modified classical Thomson
formula to atom-atom collisions. Unfortunately, as already observed
by Steenbock and Holweger, Drawin's formula allows at best an
order-of-magnitude estimate of the importance of H collisions.
Significant differences have been found compared to more reliable
estimates (e.g. experimental data by Fleck et al. 1991; calculations
for Na by Belyaev et al. 1999; for Li by Belyaev \& Barklem 2003;
and for H+H by Barklem 2007b), differing by up to six orders of
magnitude for some transitions. The uncertainty on the efficiency of
H collisions has a significant impact on non--LTE studies. There is
some evidence (Allende Prieto, Asplund \& Fabiani Bendicho 2004)
that for oxygen, relatively efficient collisions with H need to be
included in statistical equilibrium calculations. We confirm the
impact of their thermalizing effect, especially at low [Fe/H] (see
Table \ref{fabt:df54}). They tend to mitigate the large
overpopulation in the lower level of the $777$~nm triplet and thus
reduce the line opacity. The non--LTE abundance corrections at low
metallicity become considerably smaller (by $\apprge 50\%$).

In the absence of more reliable calculations, the only
possibility to have some indication of the influence of H collisions
has been to adopt the Drawin recipe. Even though likely not realistic,
this may be useful in the case of oxygen, at least (inevitably) as a
first approach, to investigate the impact of H collision
efficiency. Given available evidence pointing towards the Drawin
formula generally overestimating rates for individual transitions, the
range of S$_{\rm H}=0 - 1$ adopted here would seem reasonable for our
tests, and may be expected to give an idea of the uncertainties
involved with respect to stellar abundance estimates from the $777$~nm
\Oi\, triplet.

We also tested the effect of including the recent data by Krems,
Jamieson
\& Dalgarno (2006). Even though their estimated
rates (using formula 10 in their paper) are larger than our adopted
values, they did not have a significant impact on our results since the
transition they consider in the singlet and triplet systems
($^1$D--$^3$P) involves levels that follow LTE anyway.

\vspace{0.7cm}

\subsection{Overall non--LTE effects across the parameter space}
\label{fabss:paramspace}


\begin{flushleft}

\begin{figure}
    \begin{center}
    \includegraphics[width=8.5cm,height=6.5cm]{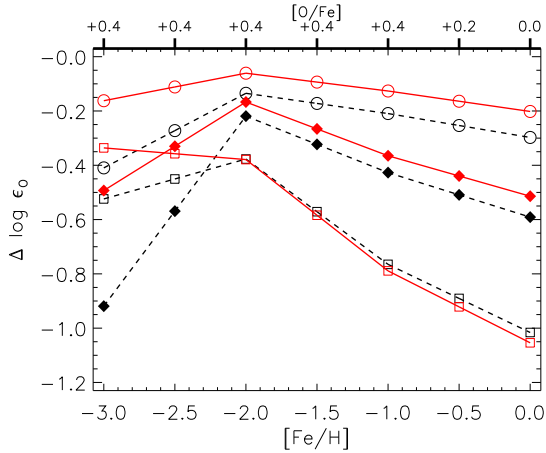}
    \caption{Non--LTE abundance corrections versus metallicity, for
    the strongest \Oi\, IR $777$~nm triplet line. The lines connect
    models with given \Teff\, and \logg\, but varying
    metallicity. Different symbols denote, respectively, results for
    models representative of a normal dwarf (\Teff$=5780$~K,
    \logg$=4.44$, indicated by open circles), a normal turn-off star
    (\Teff$=6500$~K, \logg$=4$, marked with filled diamonds) and a
    RR-Lyrae star (\Teff$=6500$~K, \logg$=2$, indicated by open
    squares).  Dashed lines are for no H collisions, solid lines for a
    choice of S$_{\rm H}=1$. Solar-scaled oxygen abundance with
    $\alpha$--enhancement below solar metallicity was adopted in the
    calculations, as indicated by the horizontal axis at the top of
    the figure\label{fabf:corr_trends}}
\end{center}
\end{figure}

\end{flushleft}


The high-excitation triplet lines are particularly sensitive to
temperature and become much stronger, both in LTE and non--LTE, for
larger \Teff\ values, because the corresponding level population
increases, i.e. there are more atoms with electrons in the proper
energy level to produce the spectral lines. Surface gravity
(pressure) controls the amount of recombination via the Saha
equation, but also the relative importance of collisional processes
with respect to radiative ones. The lines get stronger in lower
gravity models. Metal content - i.e.  smaller or larger number of
absorbers - also controls the strength of the lines.

An IDL routine\footnote{The routine and associated data are made
available for general use. They are downloadable either as
individual files, or as a single compressed file, from
ftp://nlte:c1o1nlte@ftp.iac.es (at this same address the routine for
carbon non--LTE corrections from Fabbian et al. (2006) is also
available)} has been prepared in order to interpolate among the
grid, therefore making it possible to calculate detailed abundance
corrections for any given star with parameters in the range covered.

The non--LTE line strength is larger than in LTE across the
parameter space explored, i.e., negative abundance corrections are
found. Fig. \ref{fabf:corr_trends} shows in what measure the
strongest (bluest) \Oi\, triplet line at $777.19$~nm is affected by
the resulting non--LTE corrections for some of the models (typical
dwarf, typical turn-off star, and RR Lyrae star). A qualitatively
similar trend is seen, namely of decreasing abundance corrections
for all models from solar metallicity to \feh$=-2$ and then (except
when including H collisions in the RR Lyrae case) of increasingly
severe non--LTE corrections towards very low metallicity, in
particular for turn-off stars, where the resulting
$|\Delta\log\,\epsilon_{\rm O}|$ reaches $\sim -0.9$~dex at
\feh$=-3$, when neglecting H collisions. This reduces to $\sim
-0.5$~dex if H collisions are included. For the atmospheric models
considered, non--LTE corrections are less severe around \feh$=-2$,
but still remain significant, reaching $\sim -0.4$~dex for the RR
Lyrae star. Interestingly, especially for the latter, H collisions
do not have much impact (the difference staying within $0.1$~dex for
all models) unless \feh$<-2$. At higher metallicity, including H
collisions in the RR Lyrae case actually makes the non--LTE
abundance corrections ever so slightly more severe\footnote{It is
appropriate to note that some authors have suggested to calibrate H
collisions on such stars (Clementini et al. 1995; Gratton et al.
1999). Even though RR Lyrae may be prone to very large non--LTE
effects, if adopting this strategy, it appears from our results that
it would be best to calibrate on metal-poor turn-off stars, where
the sensitivity to the choice of H collisions is larger}. For the
Sun, non--LTE corrections are $-0.3$~dex or less severe. In the RR
Lyrae case, the abundance corrections at solar metallicity are very
large and $\sim -1$~dex.


\begin{flushleft}

\begin{figure*}
    \begin{center}
    \includegraphics[width=8cm]{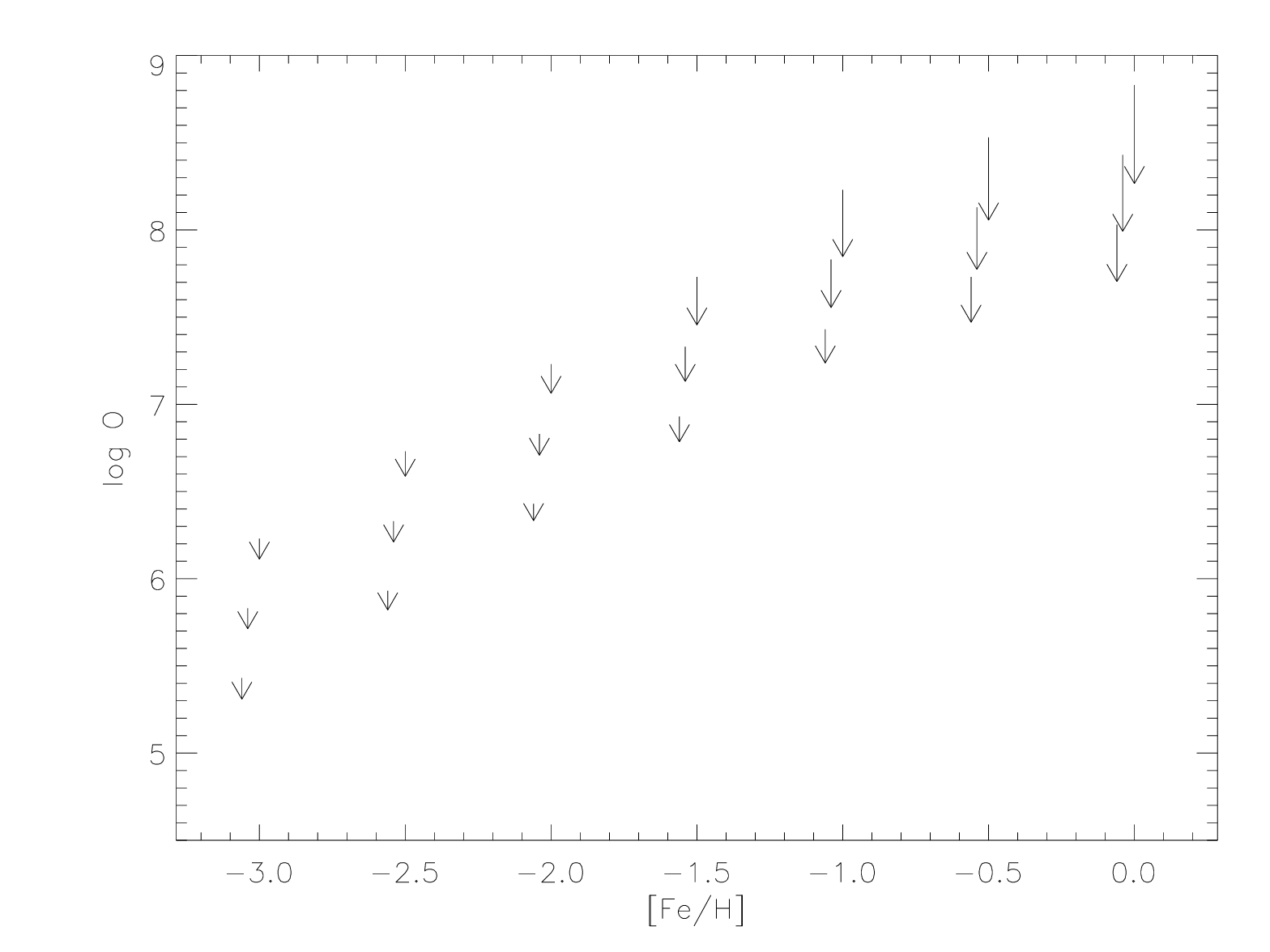}
    \includegraphics[width=8cm]{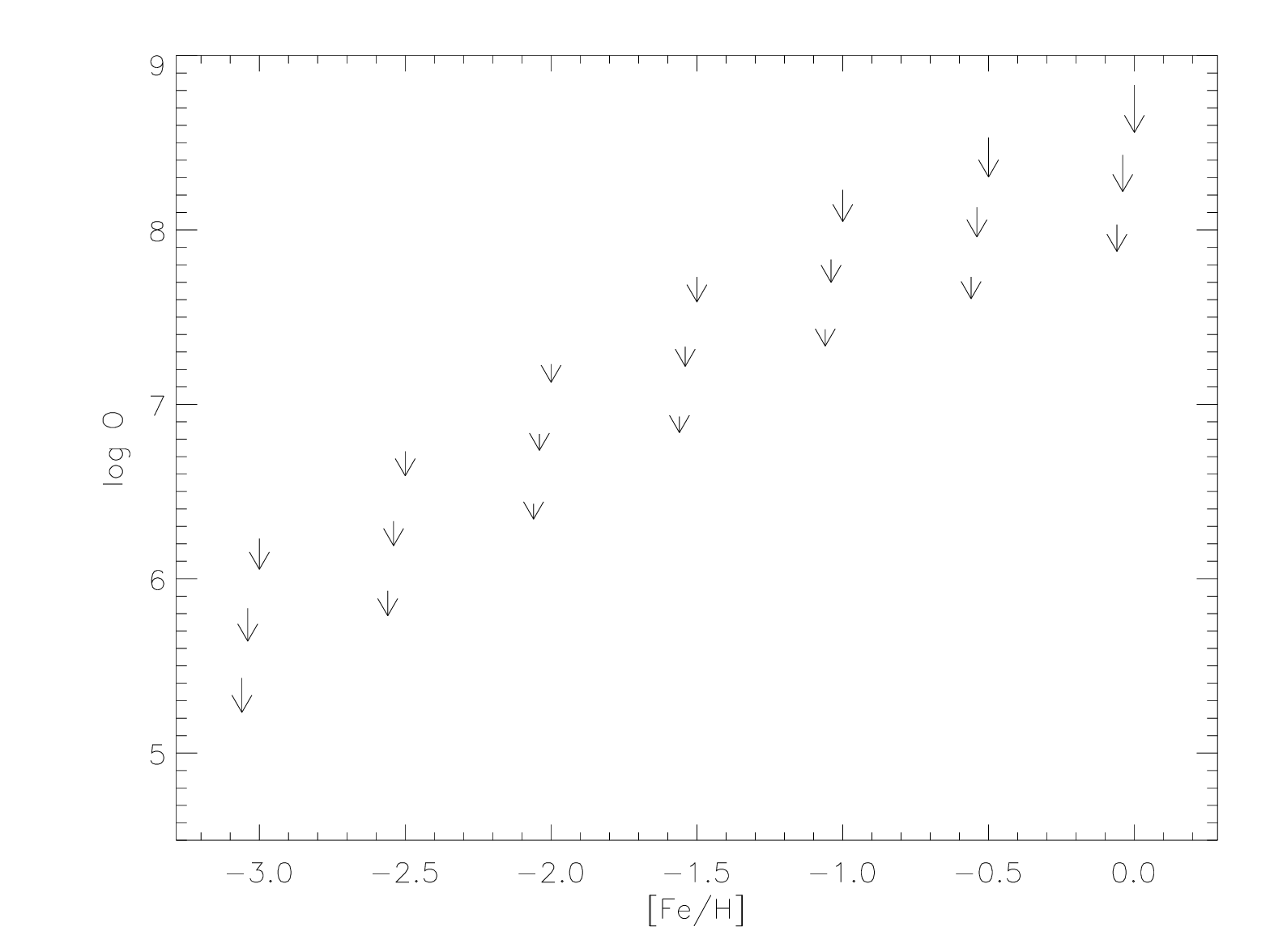}\\
    \includegraphics[width=8cm]{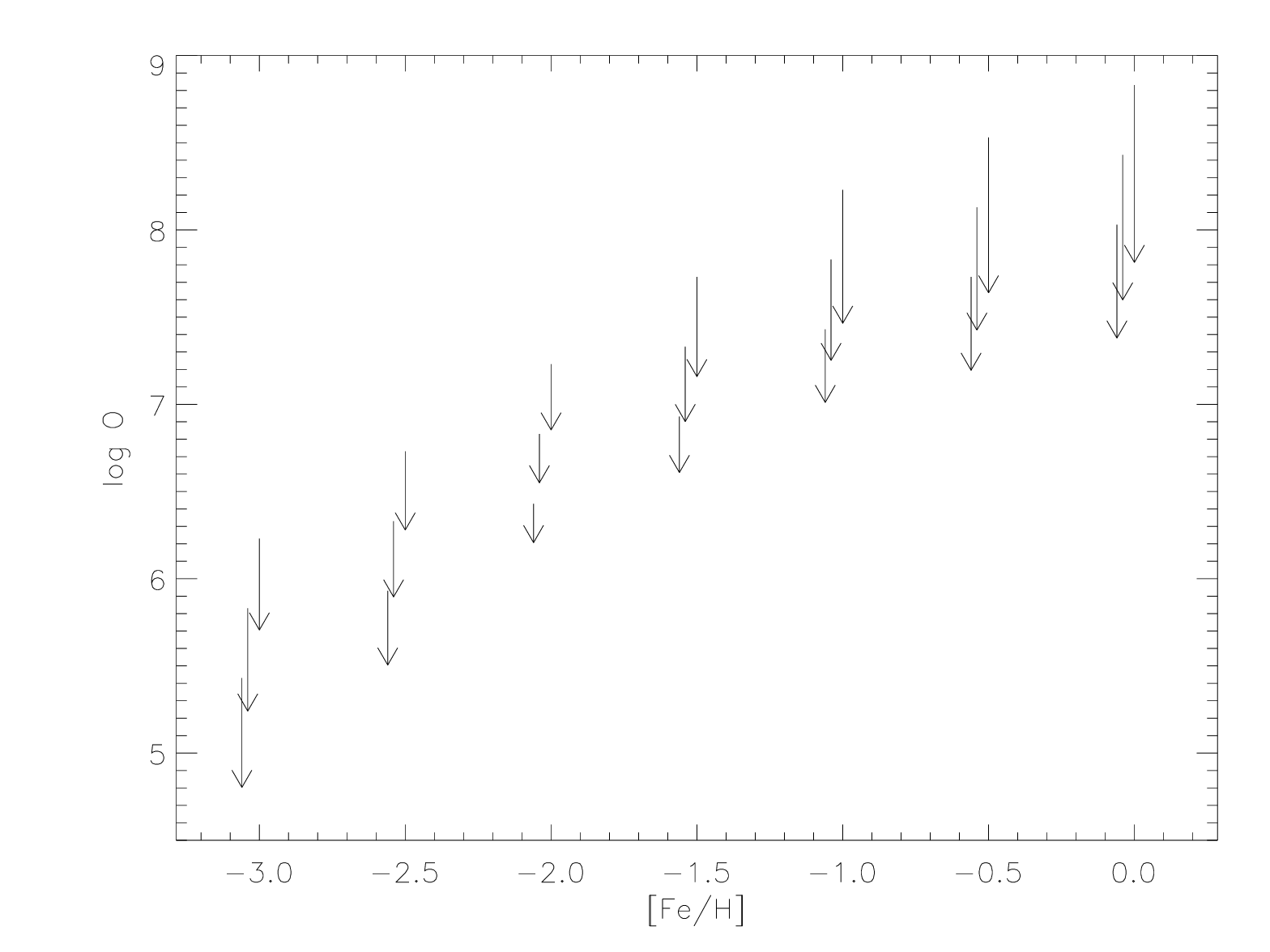}
    \includegraphics[width=8cm]{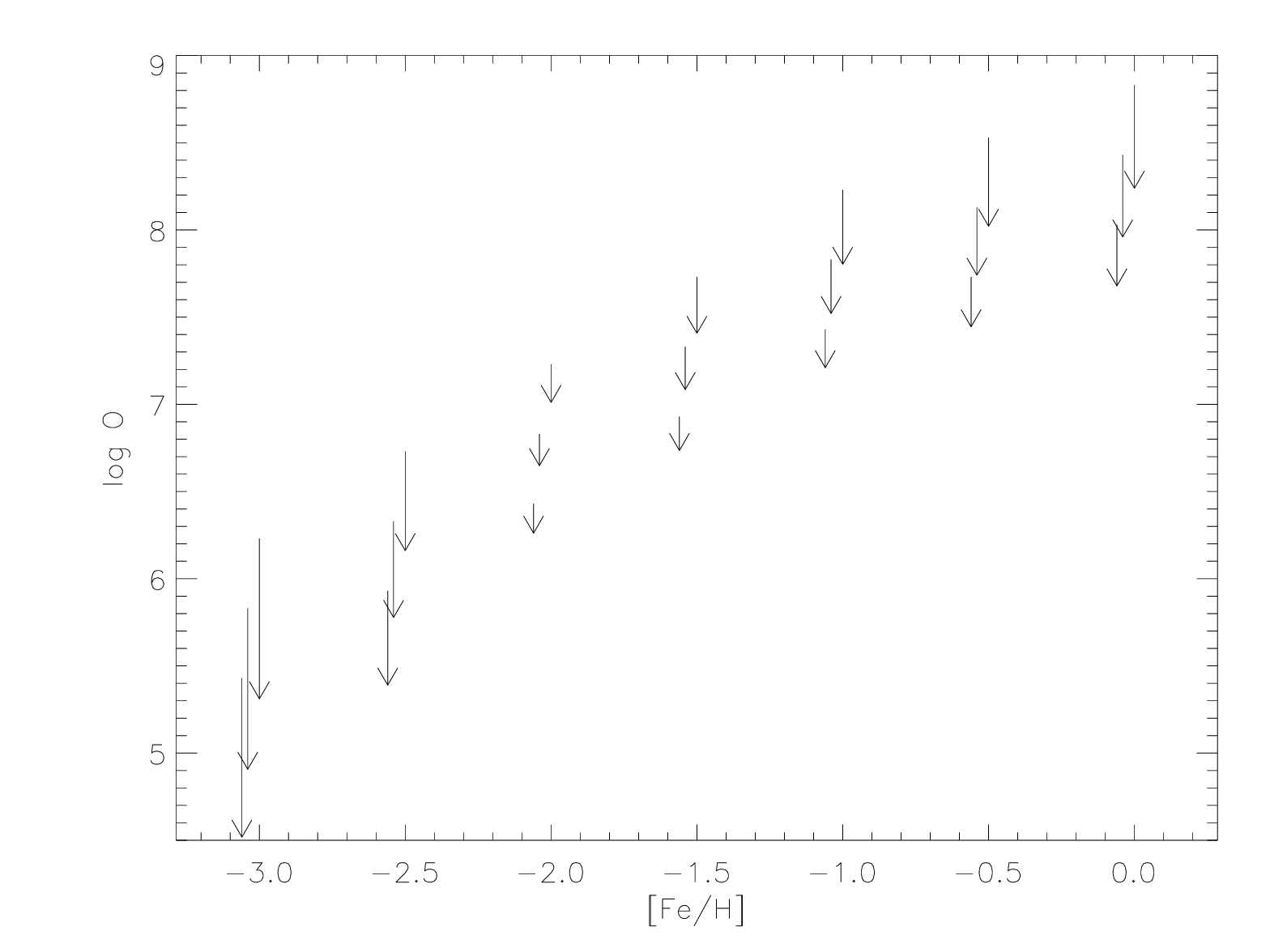}
    \caption{Non--LTE abundance corrections versus metallicity,
    for the strongest \Oi\, IR $777$~nm triplet line. {\it Top}:
    \Teff$=5500$~K; \logg$=2$ ({\it left panel}) and \logg$=4$ ({\it
    right panel}) respectively.  {\it Bottom}: same, but for
    \Teff$=6500$~K.  Results shown in all four panels are computed
    neglecting collisions with H atoms. The non--LTE corrections tend
    to get large at high temperature, both in dwarfs and giants. They
    are significant also in cooler solar-metallicity
    giants\label{fabf:logo_trends}}
\end{center}
\end{figure*}

\end{flushleft}


At metallicities $-2 < \feh\, < 0$, the line source function roughly
follows a two-level approximation and thus the non--LTE corrections
will depend on line strength. This explains their increase in
atmospheric models having higher \Teff\, and \feh, and lower \logg\,
(see Fig.\,\ref{fabf:logo_trends}).

In a more general sense (see Fig.\,\ref{fabf:logo_trends}, where we
show our resulting non--LTE corrections for a range of atmospheric
parameters across the grid explored), the abundance corrections
increase in size for higher effective temperature atmospheric models
due to increased photon losses and/or level overpopulation via
collisions. Corrections are small for \Teff$\le 5500$~K, except for
solar-metallicity giants, where they can reach $\sim -0.6$~dex. The
trend with \logg\, and \feh\, is more complex. For higher metallicity
models, the non--LTE line source function drop due to photon losses
has a two-level nature, therefore naturally increasing for stronger
lines. Thus, corrections tend to become larger with increasing
temperature and metallicity/oxygen abundance and with decreasing
\logg\, in this higher metallicity range. Below \feh$\sim -2$ however,
the influence of \logg\, and
\feh\, is the opposite. The non--LTE
corrections tend to increase strongly towards very low metallicity and
increasing gravity, suggesting that line formation is controlled by a
distinct non--LTE mechanism compared to the higher metallicity
regime.

Even though the $777$~nm triplet lines get much weaker towards low
metallicity, departures from LTE tend to increase, becoming very
large for typical metal-poor turn-off stars. An increase in gravity
at low metallicity tends to make the abundance corrections larger,
by increasing the collisional rate in the important intersystem
coupling channel. As mentioned in Sect. \ref{fabsss:UVopacity}, the
non--LTE effect is controlled by Si\,{\sc i}\, photoionization in
the continuum. Thus, when the other parameters (\Teff, \logg, $\xi$
and $\log\,\epsilon_{\rm O}$) in the atmospheric model are kept
constant, decreasing metallicity alone results in increasing
non--LTE effects because of smaller continuum opacity and generally
less efficient collisions (smaller contribution to free electron
pool). For example, even when adopting efficient H collisions
(S$_{\rm H}=1$), the abundance corrections jump from $\sim -0.5$ to
$\sim -0.85$~dex in a model with \Teff$=6500$~K, \logg$=4$ and
$\log\,\epsilon_{\rm O}=6.5$, when moving from \feh$=-3$ to $-3.5$.
Finally, oxygen abundance, while controlling the {\it strength} of
the lines, does not have a crucial influence on the abundance
corrections, with differences in $\Delta\log\,\epsilon_{\rm O}$
remaining within $\sim 0.1$~dex for a change of $0.5$~dex in
$\log\,\epsilon_{\rm O}$.

In summary, the non--LTE effects are particularly significant at the
highest temperatures, in particular for low-gravity,
solar-metallicity, and for high-gravity, very low metallicity
models. In both cases, when H collisions are neglected, the non--LTE
abundance corrections reach up to $|\Delta\log\,\epsilon_{\rm
O}|\sim 1$~dex. The inclusion of collisions with hydrogen atoms
between all energy levels generally reduces the effects
substantially (several tenths of a dex) at low metallicity, while at
higher metallicity it has a smaller impact, due to radiative losses
in the line itself being mostly important. The effects amount to
$-0.3$~dex or less in the Sun, but are $-0.4$~dex or larger for
hotter, solar-metallicity stars like Procyon.

In addition, we have tested the influence of using ATLAS9 model
atmospheres from Castelli \& Kurucz (2004) without ``convective
overshooting'', instead of MARCS models. The temperature
stratification in the MARCS and ATLAS models is generally thought to
be quite similar in this parameter space (Gustafsson et al. 2008),
so that one would expect the resulting non--LTE corrections to be
not too different in magnitude. Any uncertainty from such supposedly
small differences should be much less than that due to, say, choice
of H collision efficiency or use of the 1D approximation. We carried
out a number of test calculations which show that, when using
Castelli \& Kurucz atmospheric models describing the Sun
and stars of intermediately-low metallicity, the resulting non--LTE
abundance corrections remain in fact very similar (within a few
hundredths of a dex) to those obtained using corresponding MARCS
models. The results for the most metal-poor cases reveal that for
[Fe/H]$\le -2.5$, the non--LTE abundance corrections obtained at
higher-gravity become more significant with decreasing metallicity
also for ATLAS model atmospheres. Table \ref{fabt:kurmar} lists the
results obtained with the two sets of models. Compared to the MARCS
case, the ATLAS non--LTE corrections, although still very
significant, tend to be less dramatic. The largest discrepancy
between results using metal-poor MARCS and ATLAS models is found at
low gravity. For both sets of atmospheric models, an LTE description
of the \Oi\, IR triplet is particularly unrealistic for the
lowest-metallicity ([Fe/H]$=-3.5$) dwarf case at \Teff$=6500$~K.

Knowledge of how the two sets of models compare allows us to include
in the next section a discussion of our results in relation to other
studies which have employed ATLAS model atmospheres. Since it is
almost impossible to disentagle and remove all the different
systematics, often due to lack of sufficient information in previous
works, we have chosen to compare at face value with other non--LTE
studies in the literature, and then to apply our results directly to
relevant existing observations.

A more detailed study comparing oxygen abundance corrections
obtained with different sets of atmospheric models, should however
be carried out, in particular {\it at very low metallicity}, where
our tests reveal that when using Castelli \& Kurucz models, the
non--LTE effects are less severe compared to results obtained using
MARCS models. Within the parameter space explored for the two sets
of models, the largest differences (several tenths of a dex when
neglecting H collisions) between the \Oi\, $777$~nm triplet non--LTE
corrections appear at the lowest metallicities, both for giant and
dwarf models.

\begin{flushleft}

\begin{table}
 \centering
 \begin{tiny}
 \caption{Comparison of results obtained with representative cases from two different model atmosphere sets (ATLAS,
 and MARCS), assuming a solar oxygen abundance of $\log\,\epsilon_{\rm O}=8.66$ (Asplund et al. 2004). We list the non--LTE corrections
 (obtained using our 54-level atomic model, respectively excluding and including the contribution
 of H collisions, i.e. S$_{\rm H}=0/1$) for the central line in the \Oi\, IR triplet. The adopted $\alpha$-element abundance for all atmospheric
 models below solar metallicity is enhanced by $+0.4$~dex\label{fabt:kurmar}}
 \begin{tabular}{lccc}
        \hline\hline
       \Teff (K)$\setminus$\logg                          & 5500$\setminus$4.5 & 6500$\setminus$2.0             & 6500$\setminus$4.0  \\
\hline
                                                          &                    &                                &                     \\
                                                          & $\Delta\log\,\epsilon_{\rm O}$ & $\Delta\log\,\epsilon_{\rm O}$ & $\Delta\log\,\epsilon_{\rm O}$ \\
                                                          & {\bf (ATLAS)}      & {\bf (ATLAS)}                  & {\bf (ATLAS)}       \\
\hline
\feh$=\,0.0$                                              &    -0.20/-0.11     &      -0.83/-0.86               &   -0.53/-0.45       \\
\feh$=-1.0$                                               &    -0.14/-0.05     &      -0.64/-0.67               &   -0.36/-0.31       \\
\feh$=-2.0$                                               &    -0.11/-0.03     &      -0.32/-0.35               &   -0.20/-0.15       \\
\feh$=-2.5$                                               &    -0.12/-0.03     &      -0.28/-0.26               &   -0.27/-0.16       \\
\feh$=-3.5$                                               &    -0.43/-0.10     &      -0.26/-0.18               &   -0.87/-0.54       \\
        \hline
                                                          &                    &                                &                     \\
                                                          & $\Delta\log\,\epsilon_{\rm O}$ & $\Delta\log\,\epsilon_{\rm O}$ & $\Delta\log\,\epsilon_{\rm O}$ \\
                                                          & {\bf (MARCS)}      & {\bf (MARCS)}                  & {\bf (MARCS)}       \\
\hline
\feh$=\,0.0$                                              &    -0.18/-0.10     &      -0.88/-0.92               &    -0.50/-0.44      \\
\feh$=-1.0$                                               &    -0.12/-0.05     &      -0.63/-0.66               &    -0.34/-0.29      \\
\feh$=-2.0$                                               &    -0.11/-0.03     &      -0.30/-0.31               &    -0.19/-0.14      \\
\feh$=-2.5$                                               &    -0.16/-0.04     &      -0.42/-0.33               &    -0.56/-0.34      \\
\feh$=-3.5$                                               &    -0.52/-0.14     &      -0.88/-0.53               &    -1.21/-0.85      \\
        \hline\hline
 \end{tabular}
\end{tiny}
\end{table}

\end{flushleft}

The smaller departure coefficients in the very low-metallicity
Castelli \& Kurucz models is caused by reduced intersystem coupling
(lower rates) via electron collisions, and thus less severe atomic
level overpopulation. This is likely due to the effect of existing
differences in electron density/temperature stratification in deep
atmospheric layers (for example, up to around $400$~K at [Fe/H]$=
-2.5$), between the MARCS and ATLAS sets of very metal-poor models.
In this context, the failure of the Kurucz models to provide
matching oxygen abundances from the near-IR triplet and the
forbidden line, has previously been attributed to inconsistencies in
electron density and temperature gradients of those models near the
continuum-forming layers (Israelian et al. 2004). Compared to the
MARCS models, we find that the effect of increasingly severe
non--LTE corrections with decreasing metallicity starts to appear at
lower metallicity in the Castelli \& Kurucz models (for example,
reaching $\Delta\log\,\epsilon_{\rm O}\sim -0.7/-0.9$~dex
with/without H collisions, for a \Teff$=6500$~K, \logg$=4$ and
\feh$-4$ model). We can only speculate that the two sets of models
behave differently due to differences in opacities, and/or equation
of state, and/or convection treatment. The issue of which set of 1D
models is more suitable or realistic is still open (see discussion
in Gustafsson et al. 2008) and of course, current further
improvements in the modelling (e.g. use of three-dimensional model
atmospheres) may prove crucial.

We here warn the reader that there will be systematic errors when
applying, as is often done, corrections computed for one set of
model atmospheres, to LTE results obtained with a different set of
models, for example due to differences between continuous opacities
and ionization balances used in the construction of the model
atmospheres and those employed for the non--LTE calculations.
Therefore, the abundance corrections we derived in this paper
should, strictly speaking, be applied only to results obtained with
MARCS model atmospheres. However, given the above discussion, we
suggest that they can also be safely applied to LTE abundances from
Kurucz models down to intermediately-low metallicity.

\vspace{0.7cm}

\section{Comparison with other theoretical non--LTE studies for oxygen}

We now discuss our results in relation to those of previous work
present in the literature.

\subsection{Kiselman 1991, 1993, 2001; and related works}
\label{fabss:Kis_comparison}

Kiselman (1991, 1993, 2001) has studied and reviewed the non--LTE
effects on oxygen lines. A comparison with those works, and with
CJ93, Nissen et al. (2002), Akerman et al. (2004), and Garc\'{\i}a
P{\'e}rez et al. (2006) is of significance. Our results are also
relevant to the publications by Israelian et al. (2004), Shchukina,
Trujillo Bueno \& Asplund (2005) and Mel\'{e}ndez et al. (2006),
where the various authors used atomic data based on CJ93. In fact,
all the works mentioned in this subsection are at least in part
based on the (14- and 45-level) atomic models in that study.

The CJ93 atomic models include estimates by van Regemorter (1962)
and the impact approximation of Seaton (1962) for collisions
involving radiatively allowed bound-bound transitions, and data
available from various authors for radiatively forbidden transitions
connecting to the ground state and to the continuum. Collisional
coupling between the singlet, triplet and quintet systems is largely
absent in those models due to lack of data for several transitions,
in particular, for intersystem coupling via the radiatively
forbidden 3s $^3$S$^{\rm o}$ - 3s $^5$S$^{\rm o}$ transition. Even
though, as in CJ93, we too use the code {\small MULTI}, it is a more
recent implementation of it (version 2.3) and we have made several
other improvements in this work. Our atomic model includes a larger
number of energy levels. More crucially with respect to the $777$~nm
triplet non--LTE corrections, it features updated atomic data, and
takes into account background opacities from lines of other elements
in the region of the \Oi\, UV lines.

Kiselman (1991) studied non--LTE effects on oxygen abundances in
solar-type stars, finding that abundance corrections are slightly
decreasing from solar to \feh$=-1$ but increasingly large below
\feh$\sim -2$, reaching as much as $\sim -0.9$~dex below \feh$\sim
-2.5$ (see Fig. 3 in that work). Fig. 4 in Kiselman (1991) shows
that for a very metal-poor stellar model (\feh$\sim -3.5$~dex) the
corrections found were even more severe, with a magnitude of
$|\Delta\log\,\epsilon_{\rm O}|\apprge 1.4$~dex. These results, in
particular of non--LTE effects which gently decrease down to
intermediately low metallicity and then steeply increase at very low
metallicity, are {\it qualitatively} similar to those we obtain when
neglecting H collisions, as Kiselman did. This is somehow fortuitous
due to the opposite effect on the abundance corrections at low
[Fe/H] of, on one side, our newly adopted electron collision data,
and on the other, of both our inclusion of UV background opacity and
of the increased completeness of our atomic model.\footnote{Our
tests show collisional coupling of high-lying levels to the
continuum to be a significant line weakening channel at low
metallicity} These influences will not matter much for the Sun,
where we too find that processes in the line itself are important,
due to the line source function drop having a close to two-level
character.

Our initial tests with the atomic model by K93 made us discover that
intersystem electron collisions are crucial in the metal-poor
regime. The {\it original} model in that work gives underestimated
abundance corrections due to absence of intersystem coupling. K93
found that the introduction of intersystem collisional coupling
changed the equivalent width of the $777$\,nm triplet by a few
percent and should therefore be of minor importance. We confirm this
if testing on atmospheric models with parameters as in that study
(for Sun, Procyon and the metal-poor dwarf HD~140283). However, our
calculations on atmospheric models with both higher temperature and
lower metallicity (typical parameters of very metal-poor turn-off
stars) reveal the existence in that range of a larger influence from
intersystem coupling on the non--LTE formation of the \Oi\, IR
triplet. This is due to the fact that in these model atmospheres
there is an increased overpopulation of level 3s\,$^3$S$^{\rm o}$,
which can easily transmit to the (common) lower level of the triplet
lines due to its similar excitation energy.

For the Sun, our investigation obtains non--LTE abundance
corrections of $\sim -0.25$~dex for S$_{\rm H}=0$, in reasonable
agreement with the original results of K93. We derive a correction
of $\sim -0.15$~dex when including inelastic H collisions according
to the Drawin recipe. Compared to K93 (see also Kiselman 2001), we
find that the lower level of the $777$~nm triplet gets more
overpopulated and that its upper levels get less underpopulated.
This effect was already noted via test calculations in Kiselman's
work. It is due to atom completeness (i.e. to the combined influence
of high-excitation energy levels) caused by the significant effect
of recombination cascades due to photon losses in high-lying
infrared transitions. Coupling with the large population reservoir
in the first ionization stage exists via photon suction through
bound-bound transitions. The inclusion of high levels ensures that
the consequent efficient downward flow of electrons can be well
described. The balance between the contribution of source function
drop and line opacity will be better described by our newly employed
atomic model. We find increased contribution of line opacity. Using
our atomic model at very low metallicity (\feh$\le -3$), we find, in
particular at higher effective temperature, that coupling between
high-excitation levels and the continuum works in the opposite sense
than in the Sun, causing an upward flow.

Nissen et al. (2002) derived $777$~nm triplet non--LTE corrections
for the subset among their sample of metal-poor main sequence and
subgiant stars in which those lines had been included in the
observational setup. Using a previous version of our same code and
the model atom in K93, and neglecting H collisions, they found
moderate non--LTE corrections. The star with lowest metallicity (and
also highest temperature) in their sample is LP815-43, for which
they derive the largest non--LTE correction of
$\Delta\log\,\epsilon_{\rm O}\ = -0.25$~dex. When neglecting H
collisions as they do, we obtain $\Delta\log\,\epsilon_{\rm O}\ =
-0.55$~dex. Note however that this is likely an overestimate and
would bring the non--LTE determination of \ofe\, to very low values.
If including H collisions, the abundance correction becomes
$\Delta\log\,\epsilon_{\rm O}\sim -0.4$~dex.  It is interesting to
note that, while their non--LTE corrections show a slight decrease
with metallicity down to \feh$\sim -2.5$, there is a hint of
significant increase in the effects at or below the metallicity of
LP815-43, as stated at the end of their Section 4.3. This is the
only star falling in the parameter range where we observe the very
large increase in non--LTE corrections. Qualitatively, the trend
with metallicity in the non--LTE corrections of Nissen and
collaborators, reproduces our finding at overlapping metallicities.
Since we find collisional coupling not to play a major role until
very low metallicity is reached, this explains why only the result
for LP815-43 is significantly discrepant. The differences in the
atomic models (namely, electron and H collisions, atom completeness,
oscillator strengths) used in the two studies will indeed mostly
matter in that regime. The results of Nissen et al. were used by
Akerman et al. (2004) by interpolating the non--LTE corrections for
relevant parameters, while Garc\'{\i}a P{\'e}rez et al. (2006)
performed tailored calculations in the same fashion as Nissen et al.
The above conclusion -- namely, underestimated non--LTE effects at
very low metallicity -- applies in those cases too. For all stars in
the study by Garc\'{\i}a P{\'e}rez et al., we obtain (neglecting H
collisions as they do) more severe non--LTE abundance corrections,
however remaining within $\sim 0.05$~dex of their estimates. This
would help bring their derived \ofe\, values from the IR triplet in
better agreement with those they determined from [\Oi], however
still without reaching full consistency between the abundance
estimates from the two indicators.

Shchukina, Trujillo Bueno \& Asplund (2005) neglected H collisions
in their study of the impact of non--LTE and granulation effects on
Fe and O in the metal-poor halo subgiant HD~140283. Their resulting
1D non--LTE corrections are smaller ($-0.18$~dex) than the
correction they estimated for the Sun (which is similar to our value
for it). Such trend agrees with our result using an appropriate
model atmosphere (\Teff$=5690$~K, \logg$=3.67$ and \feh$-2.4$), for
which we too find a smaller abundance correction ($-0.12$~dex) than
our estimate for a solar model. Note that HD~140283 lies just
outside the (higher effective temperature and lower metallicity)
regime where we encounter large non--LTE effects. Moreover,
Shchukina et al.'s reanalysis of its stellar parameters, aimed at
solving the inconsistencies in the abundance determination, arrive
at the conclusion that this star has significantly lower temperature
and higher metallicity than previously thought.


\begin{flushleft}

\begin{figure}
    \begin{center}
    \includegraphics[width=8cm]{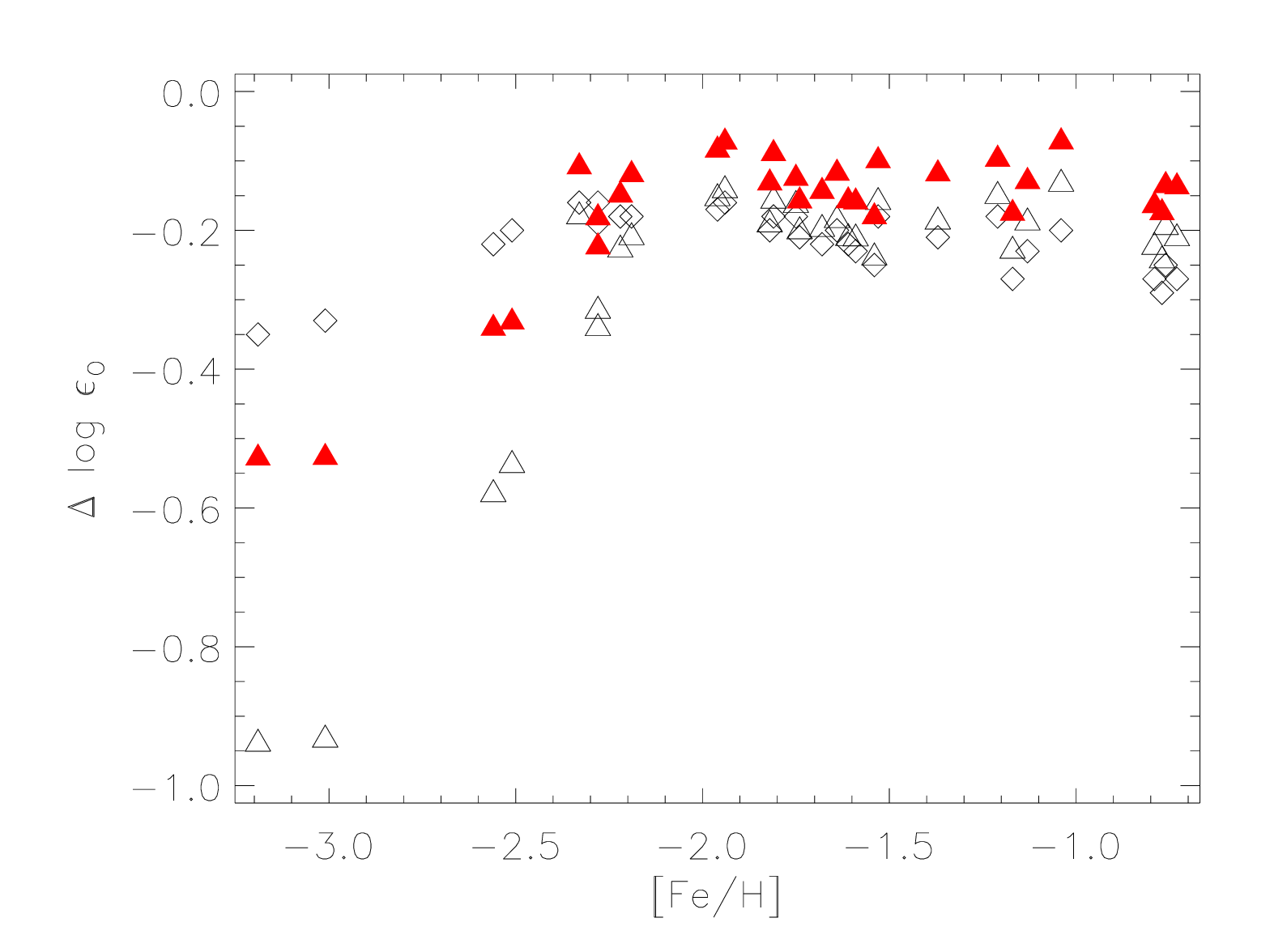}
    \caption{Non--LTE abundance corrections versus metallicity,
    for the \Oi\, IR $777$~nm triplet. Open diamonds indicate the results of
    Mel\'endez et al. (2006), while the other points mark our corrections for S$_{\rm H}=0$ and S$_{\rm H}=1$
    (open and filled triangles respectively)}\label{fabf:Meletal06}
\end{center}
\end{figure}

\end{flushleft}


Finally, regarding comparison with the work of Mel\'{e}ndez et al.
(2006), they computed non--LTE corrections with the NATAJA code (see
Shchukina \& Trujillo Bueno 2001), finding a roughly constant
oxygen-to-iron overabundance (mean \ofe$\sim +0.5$) from published
data on unevolved metal-poor halo stars. Note that their Fig. 4
shows that the lowest-metallicity stars tend to have smaller \ofe\,
values than what found around [Fe/H]$\sim -2$.  This may indicate
(see Nissen et al. 2007) too high temperatures at low [Fe/H], a
possibility either related to uncertainties in the reddening or to
the not-yet settled \Teff-scale itself. In any case, collisions with
H were not included in their work (Mel\'{e}ndez, {\it private
communication}). A comparison of the non--LTE corrections shows that
theirs are significantly smaller at low \feh. Inspecting their Fig.
5, where they compare with the non--LTE corrections of Akerman et al
(2004), the star with the largest non--LTE effect
($\Delta\log\,\epsilon_{\rm O}\sim -0.35$~dex) is G64-12, which has
the lowest metallicity in their literature sample.  For a model with
similar parameters as those used by them, we obtain a larger
abundance correction ($\sim -0.5$~dex, or more if neglecting H
collisions as they did). As seen in Fig.\,\ref{fabf:Meletal06},
while also the non--LTE corrections of Mel\'{e}ndez and
collaborators tend to become more significant at [Fe/H]$< -2$, we
find that the trend with decreasing metallicity is much more
pronounced, with more significant non--LTE abundance corrections
than found by those authors, thus likely destroying their claimed
quasi-flat [O/Fe] trend with metallicity, even when adopting the
results we obtained in the case of efficient H collisions. The
differences are likely related to the use of different codes, model
atmospheres, opacities, and to our inclusion in the oxygen model
atom of new electron-impact data.

\subsection{Gratton et al 1999}

Using an earlier version of the same code employed here and a
similar model atom to the one in K93, but Kurucz (1992) model
atmospheres, Gratton et al. (1999) studied departures from LTE in
F-K stars for a broad range of gravities and metallicities, and
including collision with electrons and hydrogen. For the latter
process, they adopted very large efficiency (scaling factor S$_{\rm
H} \sim 3.2$ in our notation), based on an empirical calibration on
RR Lyrae stars, \ie on the requirement that the $777$~nm and
$616$~nm triplets should provide the same abundances. Forcing such
agreement may however mean that other assumptions are folded into
(and remain hidden in) the scaling parameter for H collisions. Based
on previous evidence from solar investigations (Allende Prieto,
Asplund, \& Fabiani Bendicho 2004), there is some indication that
the use of Drawin's recipe with no scaling factor (i.e., S$_{\rm
H}=1$) may be reasonable for the \Oi\, IR triplet. Since collisions
with neutral H atoms are generally considered the most important
thermalizing process in late-type stars, Gratton et al.'s higher
choice for S$_{\rm H}$ will give smaller non--LTE corrections than
in our case. Indeed, those authors found relatively mild non--LTE
effects on the \Oi\, $777$\,nm triplet, in particular $\sim
-0.1$~dex in the Sun. Their study reaches down to \feh$=-3$, where
for hot, high-gravity atmospheric models they find the largest
corrections (reaching $\Delta\log\,\epsilon_{\rm O}\sim -0.5$~dex)
for the \Oi\, IR triplet. Choosing atmospheric parameters in common
between the two studies (\Teff$=6000$~K, \logg$=4.5$, \feh$=-3$, see
their Fig. 12), and very efficient H collisions as they did, we too
obtain small non--LTE abundance corrections (only $\sim -0.07$~dex
in our case, i.e. even a few hundredths of a dex less severe than in
Gratton et al.). However, if adopting S$_{\rm H} \le 1$, our
non--LTE corrections will actually be larger
($\Delta\log\,\epsilon_{\rm O}\apprle -0.25$~dex). Differences with
our study are therefore mostly due to our use of H collisions more
in line with the majority of presently available evidence, and only
to a lesser extent to employing different model atmospheres. It is
interesting to note (their Fig. 12) that their non--LTE corrections
for the hottest model used by them (\Teff$=7000$~K) show a
significant increase when going from \feh$=-2$ to $-3$. The results
of Gratton and co-workers were for example included in Carretta,
Gratton \& Sneden (2000) and in Primas et al. (2001), deriving a
``quasi-flat'' oxygen-to-iron overabundance with mean \ofe$\sim
+0.5$ at low metallicity.

\subsection{Takeda 1992, 1994, 2003; Takeda \& Honda 2005}

Takeda published a series of non--LTE studies of oxygen. In his
investigations, rates due to collisions with H atoms were in general
computed using the Drawin recipe without corrections.

Takeda (1994) found $|\Delta\log\,\epsilon_{\rm O}| < 0.1$~dex for
the Sun but $\sim -0.4$~dex for a hotter model representative of
Procyon (for which, when including H collisions, we derive a
correction of $-0.4/-0.5$~dex depending on the \Oi\, triplet line
considered). That author also attributed to non--LTE effects, at
least qualitatively, the discrepancy between oxygen abundances from
the triplet and forbidden lines generally found in studies of
low-[Fe/H] halo stars. In Takeda's non--LTE analysis, the most
oxygen-deficient (log O$=6.86$) calculation starts to show the signs
of UV pumping in the resonance lines (see their Fig. 9 and
discussion in the corresponding section).

Takeda (2003) gives an analytical formula, independent of
metallicity, to approximate the non--LTE abundance corrections
resulting from his calculations, arguing that it will provide a
fairly good approximation for lines with W$_{\lambda} < 100$~m\AA.
For the Sun (see also Takeda \& Honda 2005) and for solar-type
dwarfs at moderately low [Fe/H], our calculations including H
collisions give similar results to those of Takeda. The mentioned
formula by Takeda does not {\it explicitly} contain \feh. However,
given its dependence on line strength, it will give generally less
severe abundance corrections at low metallicity due to weaker lines,
so that our non--LTE corrections become larger than Takeda's at low
metallicity. Note that his Fig. 4 shows that estimated corrections
are low and roughly constant ($\sim -0.1$~dex) for the stars
considered in his comparison with Nissen et al. (2002). As discussed
in \ref{fabss:Kis_comparison}, we obtain roughly similar non--LTE
corrections to Nissen et al. (2002), or even more severe effects for
the lowest-metallicity star in their sample. Note however that when
adopting efficient H collisions as in the study by Takeda, the
results will be closer. The absence of a significant trend of the
non--LTE corrections with decreasing metallicity found by Takeda
(2003) is at variance with the conclusions of e.g. Carretta et al.
(2000). Takeda argues that, even though in shallow atmospheric
layers there is an appreciable increase of the non--LTE line opacity
at low metallicity, this does not affect the deeply-forming $777$~nm
\Oi\, triplet, so that fine details of the atomic model will not
matter, due to a close-to-two-level line source function. Note that
Takeda does in fact find a metallicity dependence when neglecting H
collisions. We find that at low metallicity, the two-level nature of
the non--LTE effect breaks down. In particular, for \Teff$=6500$~K,
\logg$=4$, using Takeda's formula with our estimated $W_{\rm
non-LTE}$ would give a constant $\Delta\log\,\epsilon_{\rm O}\sim
-0.1$~dex at very low metallicity, compared to our results of $\sim
-0.5$~dex and $\sim -0.9$~dex (respectively for \feh$=-3$ and
$-3.5$) when adopting H collisions. In this respect, our abundance
corrections would be more helpful in solving the oxygen discrepancy,
bringing the abundances derived from the $777$~nm triplet in better
agreement with the usually lower value from [\Oi]-based LTE
estimates.

Thus, we do not confirm the small dependence of \Oi\, non--LTE
corrections from \feh\, found by Takeda. That author (Takeda, {\it
private communication}) included intersystem collisional coupling
using Auer \& Mihalas (1973) and, in Takeda (1992) and successive
works, treatment of scattering in the continuum source function.
Note that Takeda accounted for background line opacity in the UV
only from H lines. However, as seen from our tests, metal lines tend
to make the non--LTE effect smaller. Therefore, this actually makes
the discrepancy with our results more significant, and likely
explainable due to the summed effect of different atomic data, and
use of different model atmospheres. The results in Takeda (2003)
were applied by means of interpolation/extrapolation in Takeda \&
Honda (2005), giving a linear increase of \ofe\, with \feh.

\subsection{Ram{\'\i}rez, Allende Prieto \& Lambert 2007}

In their study of a large sample of nearby stars, Ram{\'\i}rez,
Allende Prieto \& Lambert (2007) have derived non--LTE corrections
(neglecting the role of inelastic H collisions) for a grid of model
atmospheres, mainly covering the metallicity regime of the Galactic
thin and thick disks. Their adopted \Oi\ model atom is explained in
detail in Allende Prieto et al. (2003). It consists of 54 levels
plus continuum and 242 transitions, with the radiative data coming
from the Opacity Project and NIST, and electron collision rates
estimated via approximate formulae. For excitation by electrons in
forbidden transitions, they computed the collision rate coefficient
from Eissner \& Seaton (1974), adopting an arbitrary choice of
effective collision strength. As shown in the detailed comparison by
Barklem (2007a), the data here employed by us for radiatively
forbidden transitions show poor agreement with those estimated by
Allende Prieto et al. In particular, for the radiatively forbidden
transition which we find important in coupling triplet and quintet
systems, our adopted rate is two orders of magnitude larger.

The non--LTE corrections of Ram{\'\i}rez et al. reach down to $\sim
-0.4$~dex at \feh$\sim -1.5$ (their Fig. 7). For stars with
determined metallicities, as from their Table 6, the largest
corrections are $\sim -0.3$~dex (e.g., HD~112887 at \Teff$=6319$~K,
\logg$=3.88$, \feh$=-0.38$). Their abundance corrections show a
relatively mild dependence on metallicity for the range studied.
Since our investigation extends to much lower metallicity than
theirs, we only compared the results for a few representative stars
with atmospheric parameters covered by both works, using the routine
(Ram{\'\i}rez, {\it private communication}) mentioned in their
paper. The star with lowest determined metallicity in Table 6 of
Ram{\'\i}rez et al. (2007) has \feh$= -1.46$. For this star
(HD~132475), their routine gives a non--LTE correction of $\sim
-0.28$ dex with their adopted parameters of \Teff$= 5613$~K and
\logg$= 3.81$. Our non--LTE correction for similar parameters is
smaller, $\sim -0.17$ dex.

For a standard star like Procyon (\Teff$=6530$~K, \logg$=3.96$,
\feh$=0$), for which their formula would give $\sim -0.28$~dex, we
derive a much larger estimate of $\sim -0.55$~dex, when adopting
S$_{\rm H}=0$. Note however that, as discussed, a choice of S$_{\rm
H} \sim 1$ may be more appropriate, which would lead to a value of
$\Delta\log\,\epsilon_{\rm O}\sim -0.45$~dex in our case.

For the Sun, their non--LTE correction is roughly $\sim -0.15$~dex,
which is $\sim 0.1$~dex less severe than in our case. Even adopting
their slightly higher (LTE) solar oxygen abundance, our results for
the Sun would not change noticeably. A choice of S$_{\rm H} \sim 1$
for H collisions in our case would bring, however fortuitously, our
estimate close to that of Ram{\'\i}rez et al.

\section{Implications}


\begin{flushleft}

\begin{figure*}
    \begin{center}
    \includegraphics[width=9.1cm]{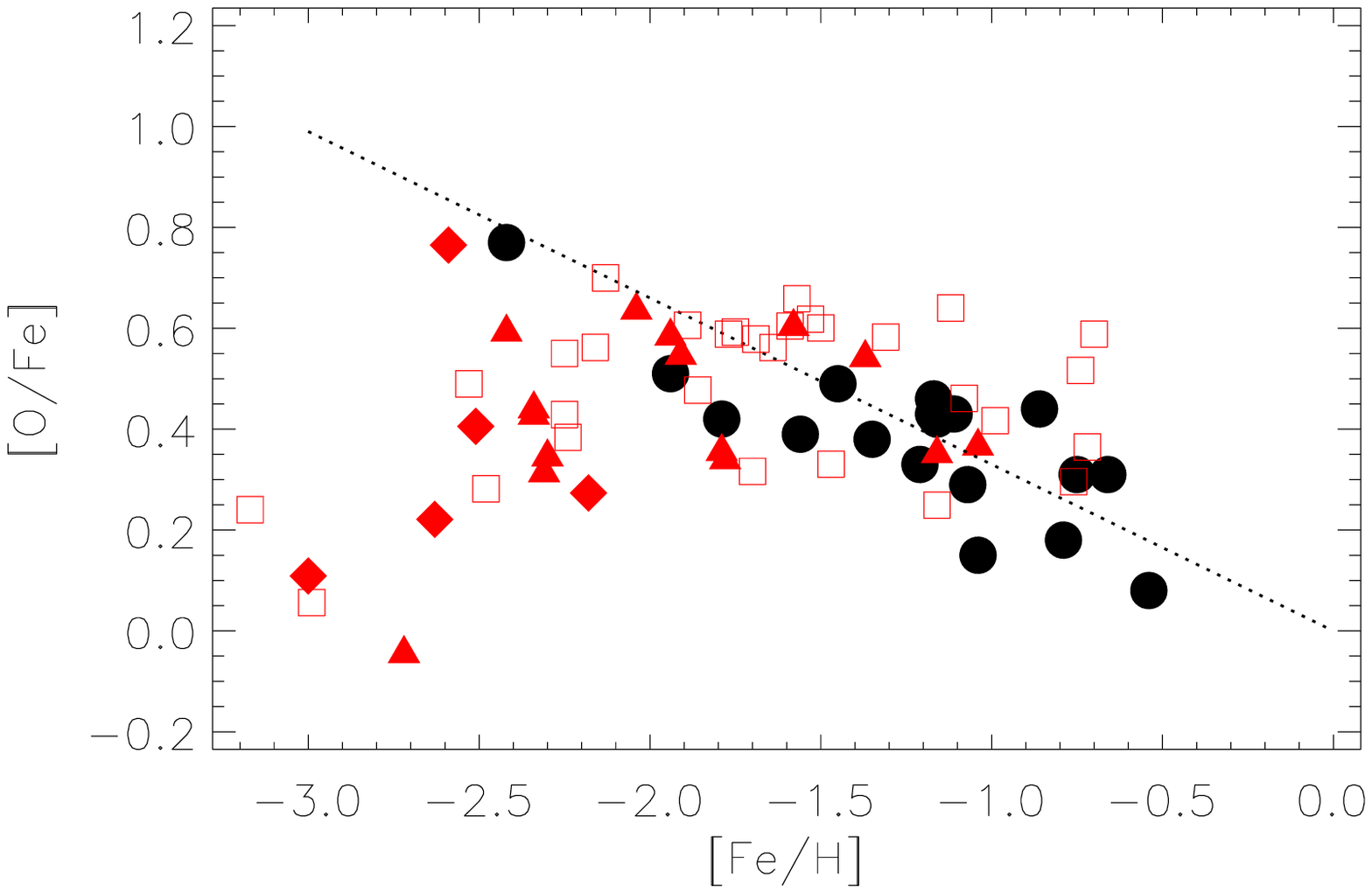}
    \includegraphics[width=9.1cm]{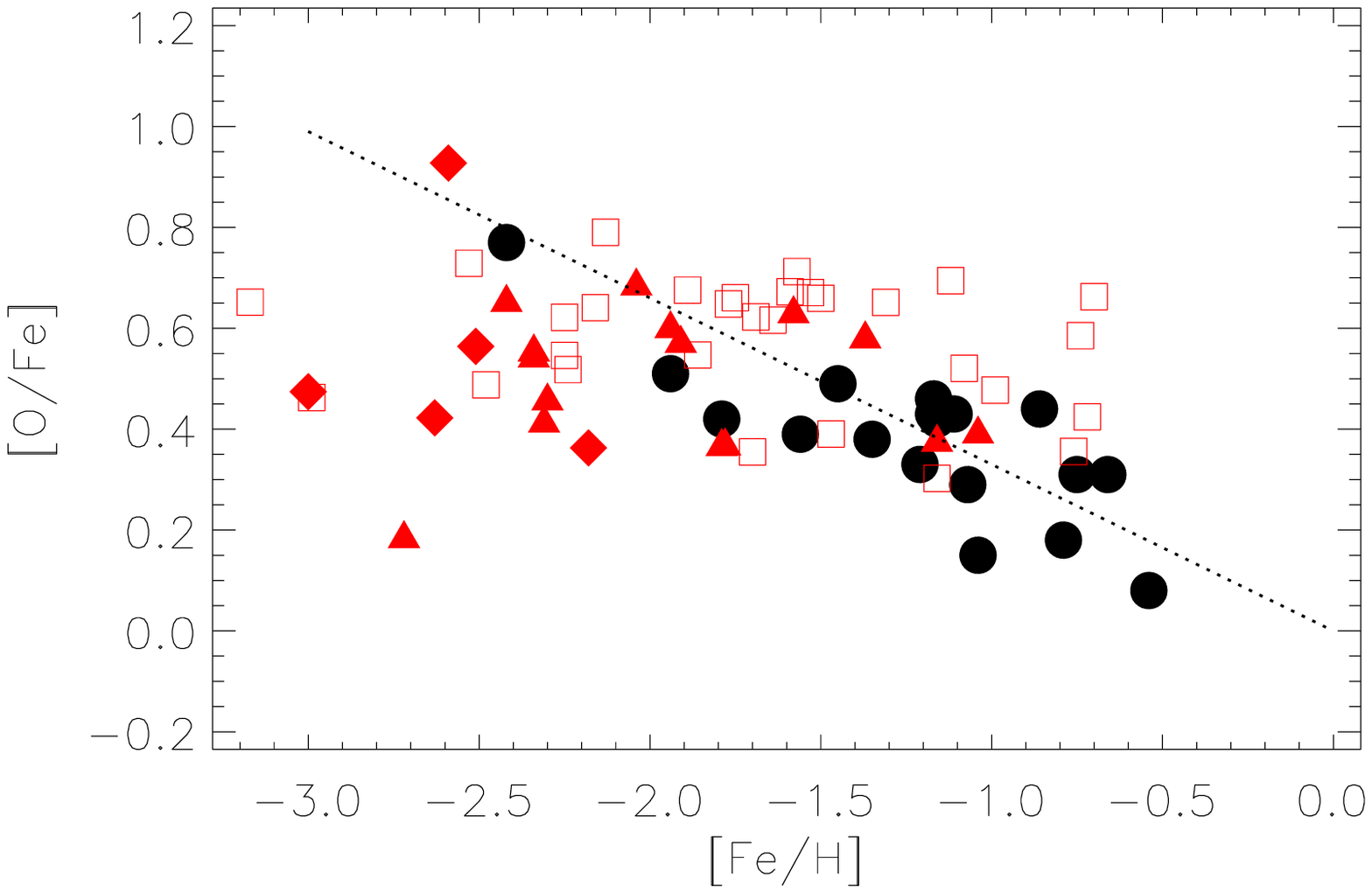}
    \caption{
    Revised [O/Fe] versus [Fe/H], from literature data. The different symbols
    are: filled circles, for [OI]-based, LTE-obeying values from
    Nissen et al. (2002); filled triangles, for the 1D estimates (from that same study,
    but non--LTE corrected according to our results) based on the \Oi\, $777$~nm triplet;
    filled diamonds for \Oi-based values for the stars in Israelian et al. (2001)
    with fully determined stellar parameters and oxygen triplet
    abundances in their study, taking into
    account our non--LTE corrections on the oxygen triplet; and empty squares, for our non--LTE corrected estimates using the
    Mel\'endez et al. (2006) data set (after adjusting \Teff-scale, see Nissen et al. 2007). The non--LTE
    abundance corrections we applied were computed with a choice of
    S$_{\rm H}=0$ and S$_{\rm H}=1$ (left and right panel respectively) for hydrogen
    collisions. The dotted line indicates the mean trend suggested by Israelian et
    al. (2001), i.e. [O/Fe]$=-0.33$\,[Fe/H]\label{fabf:ofenlte}}
\end{center}
\end{figure*}

\end{flushleft}


Our results have important consequences in relation to the debate
regarding the \ofe\, trend at low metallicity, and on the
understanding of the chemical evolution of this element in the early
Galaxy. The large difference often found in the literature between
the oxygen content of old stars as derived via the \Oi\, IR triplet
and the other oxygen abundance indicators (in particular, [\Oi]) is
likely to be explained in terms of various uncertainties acting to
produce disagreeing estimates (Asplund 2005; Mel\'endez et al.
2006).

In our re-assessment of non--LTE effects on the \Oi\, IR triplet, we
have found that at low metallicity they are more important than what
is often assumed in the literature.  In Fig.\,\ref{fabf:ofenlte}, we
show the effect of applying our abundance corrections to literature
data by Israelian et al. (2001), Nissen et al. (2002) and Mel\'endez
et al. (2006). Since uncertainties in H collision efficiency affect
the modelling of the $777$~nm \Oi\, triplet (and corresponding
non--LTE corrections) very significantly, we have plotted results
derived applying S$_{\rm H}=0$ and S$_{\rm H}=1$, respectively. The
limited observational evidence for the Sun suggests for oxygen
rather efficient thermalization via impacts with H atoms, akin to
choosing Drawin's formula without any scaling factor (S$_{\rm
H}=1$). The severe non--LTE corrections we find seem to largely
erase the linear increasing trend towards low metallicity sometimes
found in LTE. Our estimates with a choice of S$_{\rm H}=1$ suggest a
rather flat trend below [Fe/H]$=-1$, although with significant
residual scatter. Only adopting an unrealistically high amount of
thermalization would make the non--LTE effects negligible. When
instead completely ignoring H collisions, (as discussed however,
this case may be unrealistic), the [O/Fe] values in
Fig.\,\ref{fabf:ofenlte} would become even lower (the trend turning
into a decrease with decreasing metallicity). If such a model was
borne out, it would possibly require new investigations on the
efficiency of the $^{12}{\rm C}(\alpha, \gamma)^{16}{\rm O}$ nuclear
reaction and on published nucleosynthetic yields at low metallicity.

We note that our non--LTE corrections do not destroy the good
agreement between [\Oi] and \Oi\, abundances found by Nissen and
collaborators in five of their sample stars. Since those stars are
between $-2.42<\rm{[Fe/H]}<-1.04$ they do not suffer the very large
corrections (nor the large sensitivity to H collisions) we found at
very low metallicity. In fact, our resulting non--LTE corrections,
and thus estimated abundances from the $777$~nm triplet, agree with
those of Nissen et al. (2002) within $0.1$~dex for those stars.

{Our results may imply that the oxygen-to-iron overabundance does
not depend significantly on metallicity down to [Fe/H]$=-3$, with an
essentially flat trend (this may include a slight slope in either
direction)} at very low metallicity. This is in line with the
behaviour recently found via accurate abundance estimates (taking
into account non--LTE effects) for other typical $\alpha$-capture
elements (e.g. Nissen et al. 2007). In Fabbian et al. 2008b, we
further investigate this issue, together with a study of the [C/Fe]
and [C/O] ratios at low metallicity, via new observational data.

We have shown that \Oi\, LTE abundances derived from the $777$~nm
triplet are likely large overestimates, especially at low \feh, but
there is still significant uncertainty related to the choice of H
collision efficiency. Some authors have argued on astrophysical
grounds that assuming the Drawin estimate might be reasonable, at
least for the Sun. Solar center-to-limb variation in fact not only
provides evidence that the \Oi\, $777$~nm lines are formed in
non--LTE (Altrock 1968; Shchukina 1987; Kiselman 1991), but also
suggests a possible significant role of H collisions (e.g. Allende
Prieto, Asplund, \& Fabiani Bendicho 2004). Relative to electron
collisions, these should be even more important at low metallicity.
The model with S$_{\rm H}=1$ seems to predict lines in better
agreement with astrophysical expectations from other oxygen spectral
features. However, this is only an indication, since it could well
result from errors due to the Drawin estimates and to the use of 1D
atmospheric models coincidentally acting in opposite directions.
Here, we too find that H collisions may need to be taken into
account for this element, by reasoning on the otherwise unlikely
large impact of non--LTE corrections on \ofe\, at very low
metallicity, when they are neglected. At the same time, we warn that
in fact (however unavoidable in the absence of available detailed
quantum mechanical computation for relevant O+H collisions, which
are urgently needed together with constraints via solar
observations) adopting the general Drawin recipe -- possibly with
some scaling factor -- for all transitions is a rough approximation,
since different transitions likely have different sensitivity to
such collisions.

Realistic cross-sections for continuum absorption are also needed.
We have found that Si abundance plays an important role (via
bound-free absorption) in the formation of the IR oxygen triplet at
very low metallicity. This element is commonly thought to follow the
normal behaviour of other $\alpha$-elements in being overabundant
compared to Fe at low metallicity. Cayrel et al. (2004) recently
obtained an average [Si/Fe]$\sim +0.4$ (with small scatter), which
we have adopted in the calculations. However, note that very
recently Nissen \& Schuster (2008) have found evidence that halo
stars may fall into two groups, with distinct $\alpha$-enhancement.
If this significant ($\sim 0.20$~dex) separation between
``low-$\alpha$'' and ``high-$\alpha$'' halo stars is preserved down
to very low \feh\, for silicon, it may well be an important factor
in the discussion of the oxygen problem, because a lower content of
Si would imply larger non--LTE effects on the \Oi\, triplet, for an
otherwise fixed set of atmospheric parameters. Any residual
star-to-star differences in the Si content of very metal-poor halo
stars may thus play some role in the scatter observed in [O/Fe]
versus \feh. It is therefore important, in future studies on the
oxygen problem, to pin down the non--LTE corrections by also having
very accurate determinations of the stellar silicon content.

We cannot yet completely rule out the existence of a continuous
steeply increasing trend of \ofe\, with decreasing \feh. However,
its existence would require extremely efficient H collisions (e.g.
S$_{\rm H} \apprge 3$), in order to bring the abundance estimates
close to the {\it high extreme} set by the LTE expectation.

The oxygen-to-iron overabundance more likely reaches $\la +0.5$~dex
at very low metallicity, with a plateau or slight
metallicity-dependence. Further modelling and new high-quality
observations will be needed to completely settle the issue, but
models of Galactic chemical evolution constrained by the [O/Fe]
ratio derived from observations of stars at low metallicity, can now
hope to achieve a more definite conclusion on possible scenarios
required by derived stellar composition at early stages. Such low
(compared to LTE) [O/Fe] values in Galactic halo stars are in better
agreement with recent estimates (P\'equignot 2008) for
low-metallicity blue compact dwarf galaxies.

Regarding the solar oxygen abundance, it is little affected by the
new electron collisions. The determination of Asplund et al. (2004)
employed several different indicators obtaining remarkable agreement
between them, by performing a 3D non--LTE analysis. Our 1D non--LTE
results for the $777$~nm IR triplet lines agree within $\sim
0.05$~dex with those published by Asplund et al., when adopting
their choice of S$_{\rm H}=0$. We argue that, even if H collisions
were important for oxygen (Asplund and collaborators neglected them
in their calculations) - in which case our estimates would give up
to $\sim -0.1$~ dex {\it higher} oxygen abundance for some lines (in
particular, for the $844.67$~nm line) due to smaller corrections
than theirs - the overall change when averaging among all lines will
remain within $0.05$~dex. This is still significant, since the
derived solar oxygen abundance may be around $\sim 8.70$, which
would partially alleviate the current discrepancy with
helioseismology results.

Incidentally, we cannot explain the extremely large \ofe\,
overabundances derived in some cool giants at very low metallicity,
and the very different abundances derived in those objects from
permitted and forbidden oxygen lines respectively (Israelian et al.
2004), because in that range our non--LTE corrections tend to become
$\sim -0.1$~dex or less severe. To gain a better insight into the
various outstanding problems, detailed results from 3D non--LTE
investigations will be necessary.

In a series of papers, Schuler et al. (2004, 2005, 2006) have
studied solar-type open cluster dwarfs, finding much lower mean
[\Oi]-based abundances than for results based on the permitted \Oi\,
triplet. From the latter, they found a puzzling and abrupt increase
in LTE oxygen abundance derived from the high-excitation triplet for
stars with Teff $\apprle 5800$ K. We exclude that this trend could
be due for the most part to departures from LTE, since our resulting
abundance corrections tend to be relatively small in the relevant
temperature range.

\vspace{0.7cm}

\section{Conclusions}

We have explored the oxygen non--LTE line formation over a wide
parameter space.  Thanks to several improvements -- mainly, detailed
treatment of UV radiation field including opacities by background
lines, and extended model including recently available atomic data --
we derive estimated corrections at low \feh\, to LTE abundances
derived from the IR $777$~nm triplet that are generally larger than
usually adopted in the literature.
It is clear that the availability of adequate atomic data is crucial
to achieve a reliable non--LTE solution for this and other important
astrophysical problems. Using rates derived from newly available
estimates for electron-impact excitation by electrons (Barklem 2007a),
we have performed non--LTE calculations in one-dimensional stellar
atmospheric models of late-type stars extending to very low
metallicity, in order to understand the formation of the permitted
neutral atomic oxygen lines. The trends of non--LTE corrections across
the parameter space can be understood in terms of the different
non--LTE mechanisms in action: mainly source function drop at solar
metallicity, and large level overpopulation at low [Fe/H].

For solar-type dwarfs, the non--LTE corrections we found show
relatively small metallicity-dependence down to
\feh$\sim -2$. This is in line with previous findings by other
authors (e.g., Takeda 2003). However, at the higher temperatures of
turn-off stars at low \feh, the departures from LTE tend to increase
to large values when metallicity decreases, so that LTE abundances
will require increasingly large corrections. The two-level
approximation valid for the Sun breaks down and results no longer
mostly depend on processes in the line itself. The large non--LTE
corrections at low \feh\, follow from increased collisional
intersystem coupling. The (over)population of the lower level of the
$777$~nm triplet tends to increase and so does line opacity. The
formation of the $777$~nm lines becomes sensitive, at low metallicity,
to absorption processes in the UV continuum. This makes the non--LTE
corrections metallicity-dependent due to the influence of Si
absorption. The results have an impact on the derivation of the
``true'' [O/Fe] trend at low metallicity, with a roughly constant
plateau from literature data if including thermalization by collisions
with hydrogen atoms.

The problem of H collisions is obviously still very much open.
Progress is required, via theoretical quantum mechanical
calculations or via limb darkening observations (Allende Prieto,
Asplund \& Fabiani Bendicho 2004), in order to assess whether H
collisions can provide efficient thermalization. This is of high
priority also for a more conclusive determination of the solar
oxygen content as standard of reference in abundance determinations.
Some authors (e.g., Kiselman 1991; Nissen et al. 2002; Asplund et
al. 2004) have preferred to completely neglect H collisions, because
of experimental (e.g. Fleck et al. 1991) and theoretical (e.g.
Barklem, Belyaev \& Asplund 2003) evidence that Drawin's formula
gives usually too large (by orders of magnitude) estimates for
atmospheres of cool stars, at least for simple atoms like Li. We
find that even with a choice of S$_{\rm H}=1$, the non--LTE
corrections would still remain large at very low metallicity (up to
$-0.85$~dex at \feh$=-3.5$).

Even though when adopting extreme choices of the parameter
regulating the efficiency of H collisions, the difference in the
resulting $777$~nm triplet non--LTE corrections is relatively small
towards higher metallicity, it remains significant and $\sim
0.1$~dex for the Sun, in the sense of efficient collisions producing
smaller non--LTE effects (which would help to partially alleviate
the discrepancy with helioseismology, at least for oxygen).

It looks as though a combination of 3D/non--LTE effects, choice of
atmospheric models and of temperature scale, and observational
uncertainties both in derived stellar parameters and equivalent
widths measurements have conspired to prevent a clear solution of
the oxygen problem so far. It now seems like better agreement with
the abundance from the forbidden [O\,{\sc i}] lines (which should be
free from non--LTE effects) and from OH molecular lines (once
corrected for 3D effects, Asplund \& Garc\'{\i}a P{\'e}rez 2001) can
be obtained when non--LTE effects are taken into account for the
$777$\,nm oxygen triplet abundance, see e.g.~\/ Mel\'{e}ndez et al.
(2006). Our results seem to suggest that large non--LTE corrections
at low metallicity are the norm, at least for turn-off halo stars.

Given our findings concerning the importance of the radiatively
forbidden 3s $^3$S$^{\rm o}$ - 3s $^5$S$^{\rm o}$ transition, it is
crucial to check if a similar behaviour is seen for other atoms with
similar structure, e.g. C, N, and S, or even in all atoms.

Non--LTE corrections obtained with ATLAS models turned out to be
very similar, for a range of atmospheric parameters, to those we
found using MARCS models. However, at very low metallicity,
significant discrepancies appear, with large differences between
non-LTE corrections using the two sets of models. Interestingly,
this metallicity range overlaps with that for which the debate
between ``flat'' and ``linear'' [O/Fe] trend exists. However, as
seen from Fig.\,\ref{fabf:ofenlte}, our main conclusions will not
change even if applying ATLAS non-LTE corrections, which would give,
around [Fe/H]$=-2.5$, smaller corrections by $\sim 0.3$~dex in the
left panel (no H collisions) and by $\sim 0.2$~dex in the right
panel (H collisions 'a la Drawin'). Anyhow, more sophisticated
modelling of stellar atmospheres is likely another crucial factor
with respect to the oxygen problem, in particular for abundances
from OH lines, but also to understand how the large non--LTE effects
we found act in a 3D atmosphere. Therefore, it is obviously urgent
to carry out, for a large range of stellar parameters, full 3D
non--LTE calculations, in order to further investigate the formation
of the oxygen lines at low \feh. It is expected that non--LTE
effects will be enhanced when using 3D atmospheric models, due to
their cooler superficial temperatures, to which the high-excitation
lines of interest are known to be sensitive. This crucial step
forward will allow enhanced abundance analyses and will help to shed
light on early Galactic chemical evolution, contributions by SN II
and massive stars, time delay of yields from Type Ia SNe, and the
reality of the [C/O] upturn at low metallicity (Akerman et al. 2004;
Spite et al. 2005; Fabbian et al. 2008b). It is likely that -
together with the use of an accurate temperature scale - when
finally taking into account, using detailed calculations with the
best available atomic data, the interplay of 3D and non--LTE effects
in the formation of the various available abundance indicators
(including for the determination of metallicity), it will be
possible to in the end fully solve the ``oxygen problem''.

Here, we have demonstrated that non--LTE corrections play an
important role towards a solution. They will need to be taken into
consideration in order to abate large systematic errors still
afflicting LTE-based estimates. To check whether agreement between
abundances from the different oxygen lines can be found at low
metallicity will furthermore require using improved observational
data, in particular from a large sample of subgiants or giants where
more oxygen indicators are available. This important test will help
to clarify outstanding issues related to the [O/Fe] controversy.

\bigskip
\begin{acknowledgements}
This work has been partly funded by the Australian Research Council
(grants DP0342613 and DP0558836). DF is grateful to the Department of
Astronomy and Space Physics, Uppsala Astronomical Observatory,
Uppsala, Sweden, for its hospitality. DF would also like to thank Remo
Collet, Jorge Mel\'{e}ndez and Poul Erik Nissen for fruitful
discussion and comments. PB is a Royal Swedish Academy of
Sciences Research Fellow supported by a grant from the Knut and Alice
Wallenberg Foundation. PB also acknowledges the support of the Swedish
Research Council. This research has made use of NASA's Astrophysics
Data System, of the NIST Atomic Spectra Database (version 3),
which is operated by the National Institute of Standards and
Technology, and of the Vienna Atomic Line Database. We are indebted to Iv\'{a}n
Ram{\'\i}rez for providing the routine which we used to compare with
his non--LTE oxygen results, and to Yoichi Takeda, for details on his non--LTE work.

\end{acknowledgements}

\end{document}